\documentclass[12pt,thmsa,notitlepage]{article}
\usepackage{amsmath}
\usepackage{amssymb}
\usepackage{amsfonts}
\usepackage{sw20lart}
\usepackage[onehalfspacing]{setspace}
\usepackage{graphicx}

\input{tcilatex}

\begin{document}

\title{Superfield Calculation of Loop Contribution in Extra Dimensional
Theories }
\author{{\large Minh Q. Truong\thanks{%
minh\_t@hotmail.com}} \\
{\small Department of Physics and Astronomy,} {\small University of Missouri
- St. Louis }\\
St. Louis, MO 63121}
\maketitle

\begin{abstract}
Superfields provide a compact description of supersymmetry representations.
Loop corrections with superfield formalism are simpler and much more
manageable than calculation in terms of component fields. In this paper we
calculate the contribution of the Kaluza-Klein states, associated with extra
dimensions, to the renormalization group beta function. These Kaluza-Klein
particles circulate in the virtual loop, hence affecting the overall
corrections at any order. We obtain the one-loop correction, which checks
with the result previously obtained using the more laborious component field
method. In addition, we calculate the two-loop correction coming from chiral
KK states.
\end{abstract}

\section{Introduction}

Supersymmetry alleviates the hierarchy problem where the cancellation of the
quadratic divergences occurs in loop-calculations at all orders of
perturbation. This was first notice in the self-interacting model\cite{WZ}.
In modern language, for N=1 supersymmetry, the softening of quantum
fluctuation is called non-renormalization of the superpotential. However,
these results can be obtained more easily in the context of superfield
calculations \cite{DM-Superfield}. Superfields provide a compact description
of supersymmetric representation and a useful theoretical tool for
formulating the Lagrangians. In addition, superfields simplify the addition
and multiplication of representations. This profound concept of superfield
was proposed by \cite{SS-superfield} soon after the discovery of
supersymmetry. Superfields are just functions on superspace, where ordinary
space-time is enlarged to include fermionic coordinates. The quantum
properties of supersymmetric field theories are best investigated using
superfield perturbation theory or supergraph method. The advantage for the
supergraph method is its calculational simplicity due to fewer Lorentz
indices and simpler Dirac structure, as compare to the traditional component
field approach. \ Also, the results are manifestly supersymmetric at all
stages of calculation. The component content or fields can always be
recovered by power series expansion in terms of the anti-commuting
coordinates.

The study of superstring theory has revived interest in theories with extra
dimensions\cite{Antoniadis}\cite{HW-model}\cite{AGW-extrdim}\cite%
{MSS-extrdim}. This interest has been further stimulated in the recent years
by the possibility that these hidden dimensions may be "large" --- large
enough that their phenomenological implications can be checked by
experiments in the near future\cite{Antoniadis I.}\cite{ADD-model}. The
dependency of gauge unification scale on the size of the compactified
dimensions is such that larger compactified dimensions lead to lower
unification scale\cite{DDG-intscale}.\ In this paper I apply this superfield
Feynman diagram method \cite{GSR-supergraph}\cite{Susybooks} to loop
calculations in supersymmetry theory with extra dimensions. There are three
basic standard approaches to perturbative calculations in higher dimensional
theories: 1) by summation over the winding numbers using the Poisson
resummation formula\cite{resummation}, 2) by using mixed propagators, where
the coordinates of the uncompactified dimensions are Fourier transformed
into momentum space while the compactified dimensions are kept in
configuration space\cite{mixedpropagator}\cite{KKsum} . 3) by summation over
the Kaluza-Klein states, which are manifestation of fields confined in the
compactified extra dimensions \cite{KKtower}\cite{DDG-intscale}. Each
particle that can propagate in extra dimensions shows up as an infinite
tower of particles with masses $\frac{n}{R}$ where $n\epsilon Z$ and the
mass gaps is controlled by the size of the compactified scale.

In this paper, we follow method 3). Our model is a 4D super Yang-Mills
theory with manifestation of extra dimensions as an infinite tower of
Kaluza-Klein states. In this theory, non-chiral fields(Higgs,gauge and
scalar bosons fields) can propagate freely in all dimensions therefore will
have KK modes. Matter fields (Leptons, quarks,...), on the other hand, are
restricted to the brane. Collectively,they are called the bulk and brane
fields respectively. These Kaluza-Klein particles associated with the bulk
fields are allowed to circulate in the virtual loop, hence affecting the
overall corrections at any order. We obtained the one-loop correction with
the inclusion of KK states, and as application we computed the Beta
functions for the gauge coupling. In addition, we calculated the two-loop
partial correction with KK states where pure vector contributions are not
considered.

\section{Kaluza-Klein Contribution at One-Loop}

In this section, we consider 4D super Yang-Mills theory with manifestation
of extra dimensions as an infinite tower of Kaluza-Klein states.
Equivalently, a higher dimensional Super Yang-Mill theory where the extra
dimensions are compactified on an orbifold. In this theory, gauge fields can
propagate freely in all dimensions, but matter fields are confined to the
branes or orbifold fixed points. This higher dimensional theory is
non-renormalizable due to the increase in spacetime dimensionality. However,
we can safely assume higher Kaluza-Klein excitations are decoupled from the
theory at a given energy scale, giving rise to an approximate renormalized
theory. We will assume that there exist $\delta\equiv D-4$ extra spacetime
dimensions with compactified radius $R$, where $\mu_{0}\equiv R^{-1}$
represents the energy scale of the compactified dimensions. Every non-chiral
particle state in the MSSM with mass $m_{0}$ can have an infinite tower of
Kaluza-Klein states with masses 
\begin{equation}
m_{n}^{2}\equiv m_{0}^{2}+\sum_{i=1}^{\delta}\frac{n_{i}^{2}}{R^{2}},
\end{equation}
where each state mirrors the MSSM ground state and $n_{i}\in Z$ are the
Kaluza-Klein excitation numbers.

Before embarking on the superfield calculation, we need to know exactly
which objects we are calculating. Analogous to the self energy or vacuum
polarization in component field approach, $\overline{\Gamma},$ called
"quantum correction to the effective action", is the object which we need in
our superfield calculation. The counter terms in the Lagrangian and
subsequently the renormalization constants . To have a better understanding
of $\overline{\Gamma},$ we write the generating functional for the Green's
functions as%
\begin{equation}
Z[J]=e^{\left( \frac{iW[J]}{\hbar}\right) }=\int D\varphi e^{\frac{i}{\hbar }%
\left( S[\varphi]+\int dx\varphi(x)J(x)\right) }
\end{equation}
where $W[J]$ is the generating functional for the connected Green's
functions and the classical action $S[\varphi]$, with $\pi(x)$ being the
conjugate to $\varphi(x)$.The effective action is defined by the Legendre
transformation%
\begin{equation}
\Gamma\lbrack\pi]=W[J]-\int dx\pi(x)J(x),
\end{equation}
is the central object of quantum field theories since it contains all
necessary information about the theories. The effective action is also known
as the generating functional for the n-point vertex function%
\begin{equation}
\Gamma^{n}=\frac{\delta^{n}\Gamma}{\delta\pi\left( z_{1}\right) \cdot
\cdot\cdot\delta\pi\left( z_{n}\right) }\mid_{J=0}.
\end{equation}
The n-point vertex function represents the one-particle irreducible Feynman
diagram without the external lines. The generating functional for the
Green's functions becomes 
\begin{equation*}
e^{\frac{i}{\hbar}\left( \Gamma\lbrack\pi]+\int dx\pi(x)J(x)\right) }=\int
D\varphi e^{\frac{i}{\hbar}\left( S[\varphi]+\int dx\varphi(x)J(x)\right) }.
\end{equation*}
We make a change of variable $\varphi\rightarrow\varphi+\pi$ , the above
becomes%
\begin{equation*}
e^{\frac{i}{\hbar}\left( \Gamma\lbrack\pi]+\int dx\pi(x)J(x)\right) }=\int
D\varphi e^{\frac{i}{\hbar}\left( S[\varphi+\pi]+\int dx(\varphi
(x)+\pi(x))J(x)\right) }
\end{equation*}%
\begin{equation*}
e^{\frac{i}{\hbar}\left( \Gamma\lbrack\pi]\right) }=\int D\varphi e^{\frac{i%
}{\hbar}\left( S[\varphi+\pi]+\int dx\varphi(x)J(x)\right) }
\end{equation*}
since $\frac{\delta\Gamma\lbrack\pi]}{\delta\pi(x)}=-\int
dx^{\prime}\delta(x-x^{\prime})J(x^{\prime})=-J(x).$ 
\begin{equation}
e^{\frac{i}{\hbar}\left( \Gamma\lbrack\pi]\right) }=\int D\varphi e^{\frac{i%
}{\hbar}\left( S[\varphi+\pi]-\int dx\varphi(x)\frac{\delta \Gamma\lbrack\pi]%
}{\delta\pi(x)}\right) }.
\end{equation}
Next we Taylor expand $S[\varphi+\pi]$ in $\pi$%
\begin{equation}
S[\varphi+\pi]=S[\pi]+\sum_{n=1}^{\infty}\frac{1}{n!}\int dx_{1}...dx_{n}%
\frac{\delta^{n}S[\pi]}{\delta\pi(x_{1})...\delta\pi(x_{n})}\varphi
(x_{1})...\varphi(x_{n}),
\end{equation}
and plug this expansion into the above%
\begin{equation*}
e^{\frac{i}{\hbar}\left( \Gamma\lbrack\pi]\right) }=\int D\varphi e^{\frac{i%
}{\hbar}\left( S[\pi]+\sum_{n=1}^{\infty}\frac{1}{n!}\int dx_{1}...dx_{n}%
\frac{\delta^{n}S[\pi]\varphi(x_{1})...\varphi(x_{n})}{\delta\pi(x_{1})...%
\delta\pi(x_{n})}-\int dx(\varphi(x)\frac{\delta \Gamma\lbrack\pi]}{%
\delta\pi(x)})\right) },
\end{equation*}
which reduces to 
\begin{equation*}
e^{\frac{i}{\hbar}\left( \Gamma\lbrack\pi]-S[\pi]\right) }=\int D\varphi e^{%
\frac{i}{\hbar}\left( \sum_{n=1}^{\infty}\frac{1}{n!}\int dx_{1}...dx_{n}%
\frac{\delta^{n}S[\pi]\varphi(x_{1})...\varphi(x_{n})}{\delta
\pi(x_{1})...\delta\pi(x_{n})}-\int dx(\varphi(x)\frac{\delta\Gamma\lbrack
\pi]}{\delta\pi(x)})\right) }.
\end{equation*}
Taking the natural logarithm on both sides,%
\begin{equation*}
\Gamma\lbrack\pi]-S[\pi]=\int D\varphi\left( 
\begin{array}{c}
\sum_{n=1}^{\infty}\frac{1}{n!}\int dx_{1}...dx_{n}\frac{\delta^{n}S[\pi]%
\varphi(x_{1})...\varphi(x_{n})}{\delta\pi(x_{1})...\delta\pi(x_{n})} \\ 
-\int dx(\varphi(x)\frac{\delta\Gamma\lbrack\pi]}{\delta\pi(x)})%
\end{array}
\right)
\end{equation*}
by inspection we have%
\begin{equation}
\Gamma\lbrack\pi]-S[\pi]=\overline{\Gamma}[\pi].
\end{equation}
$\overline{\Gamma}[\pi]$ represents the quantum correction to the classical
action $S[\pi].$ The effective action $\Gamma\lbrack\pi]$ is the classical
action plus all quantum corrections.

Our calculation is similar to the method used in reference \cite%
{DDG-intscale}. We use the supergraph technique in calculating the chiral
correction of the vector superfield where computation are being carried out
in a larger space called superspace coordinates. Superspace contains the
ordinary space-time coordinates $x^{\mu }$ and four additional anticommuting
numbers $\left\{ \theta _{\alpha }\right\} _{\alpha =1,2}$ and$\left\{ 
\overline{\theta }_{\overset{\cdot }{\alpha }}\right\} _{\overset{\cdot }{%
\alpha }=\overset{\cdot }{1},\overset{\cdot }{2}}$. Superfields are then
defined to be functions of superspace. According to our model, chiral
superfields are confined to the brane, being prohibited to propagate in the
extra dimensions, will not have KK excitations. Therefore, no KK
contribution coming from the chiral superfields. The effects of the extra
dimensions through the manifestation of KK excitations can only come from
the radiative correction of the vector superfield to itself. With the
extensive use of the super Feynman rules derived in reference \cite%
{GSR-supergraph}, the chiral correction of the vector superfield takes on
the form in configuration space 
\begin{align}
\overline{\Gamma }_{\Phi }(V)& =\frac{1}{2}\int
d^{8}z_{1}d^{8}z_{2}V^{a}(x_{1},\theta _{1},\overline{\theta }_{1})  \notag
\\
& \times (igT_{i}^{aj})(igT_{ji}^{b})V_{b}(x_{2},\theta _{2},\overline{%
\theta }_{2})[D^{2}(z_{1})\frac{i\delta ^{8}(z_{1}-z_{2})}{16(\partial
_{1}^{2}-m^{2})}\overleftarrow{\overline{D}^{2}}(z_{2})]  \notag \\
& \times \lbrack D^{2}(z_{2})\frac{i\delta ^{8}(z_{2}-z_{1})}{16(\partial
_{2}^{2}-m^{2})}\overleftarrow{\overline{D}^{2}}(z_{1})]
\end{align}%
where $V^{a}$ is a vector superfield, the $z^{\prime }s$ are the
super-coordinates%
\begin{equation}
z^{M}=(x^{\mu },\theta ^{\alpha },\overline{\theta }_{\overset{\cdot }{%
\alpha }})
\end{equation}%
$T^{a}$ is the generator of the Lie group $G$ which span the Lie algebra
under the commutator%
\begin{equation}
\left[ T^{a},T^{b}\right] =if^{abc}T^{c}.
\end{equation}%
The $D^{\prime }s$ are the super spinorial derivatives and $g$ is the
coupling constant between chiral and vector superfields. The super-diagram
in momentum space corresponds to figure 1.

Using covariant algebra, the correction becomes%
\begin{align}
\overline{\Gamma }_{\Phi }(V)& =\frac{1}{2}\sum_{A}\delta
^{ab}T_{A}(R)g^{2}\int d^{8}z_{1}d^{8}z_{2}V^{a}(z_{1},\theta _{1},\overline{%
\theta }_{1})  \notag \\
& \times V_{b}(z_{2},\theta _{2},\overline{\theta }_{2})[D^{2}(z_{1})%
\overline{D}^{2}(z_{1})\frac{\delta ^{8}(z_{1}-z_{2})}{16(\partial
_{1}^{2}-m_{n_{i}}^{2})}]  \notag \\
& \times \lbrack \overline{D}^{2}(z_{1})D^{2}(z_{1})\frac{\delta
^{8}(z_{1}-z_{2})}{16(\partial _{2}^{2}-m_{n_{i}}^{2})}].
\end{align}%
where $\sum_{A}\delta ^{ab}T_{A}(R)=T_{i}^{aj}T_{ji}^{b}$ depends on the
representation $R$ on which the fields are chosen. Next, we integrate by
parts on $z_{1},$ and using the properties of the super covariant
derivatives, the correction becomes 
\begin{align}
\overline{\Gamma }_{\Phi }(V)& =\frac{1}{2}\sum_{A}T_{A}(R)g^{2}\int
d^{4}x_{1}d^{4}\theta _{1}d^{4}x_{2}d^{4}\theta _{2}V^{a}(z_{2},\theta _{2},%
\overline{\theta }_{2})\ \delta ^{8}(z_{1}-z_{2})  \notag \\
& \times \left\{ 
\begin{array}{c}
16^{2}\square \int \frac{d^{4}q}{(2\pi )^{4}}e^{iq(x_{1}-x_{2})}+16^{2}i%
\partial _{\overset{\cdot }{\sigma }}^{\alpha }\frac{1}{2}\int \frac{d^{4}q}{%
(2\pi )^{4}}e^{iq(x_{1}-x_{2})}\overline{D}^{\overset{\cdot }{\sigma }%
}D_{\alpha } \\ 
+16\ \delta ^{4}(x_{1}-x_{2})\overline{D}^{2}D^{2}%
\end{array}%
\right\}  \notag \\
& \times \frac{V^{a}(z_{1},\theta _{1},\overline{\theta }_{1})}{16(\partial
_{1}^{2}-m_{n_{i}}^{2})16(\partial _{2}^{2}-m_{n_{i}}^{2})}.
\end{align}%
Fourier transforms the vector superfield $V$ and all propagators to momentum
space yield quantities such as $p,p^{\prime },q$ and $k$ in the correction.
Then we perform the following integrations:$\int d^{4}\theta _{2},$ $\int
d^{4}x_{1},$ $\int d^{4}x_{2},$ $\int d^{4}p^{\prime },$ $\int d^{4}q$ $.$
The correction reduces to 
\begin{align}
\overline{\Gamma }_{\Phi }(V)& =\frac{1}{2}\sum_{A}T_{A}(R)g^{2}\int
d^{4}\theta _{1}\frac{d^{4}p}{(2\pi )^{4}}\widetilde{V}(p,\theta _{1},%
\overline{\theta }_{1}) \\
& \times \left\{ \int \frac{d^{d}k}{(2\pi )^{d}}\frac{[-\left( k-p\right)
^{2}+\frac{1}{2}\left( k-p\right) _{\overset{\cdot }{\beta }}^{\alpha }%
\overline{D}^{\overset{\cdot }{\beta }}D_{\alpha }+\frac{1}{16}\overline{D}%
^{2}D^{2}]}{[k^{2}+m_{n_{i}}^{2}][(k-p)^{2}+m_{n_{i}}^{2}]}\widetilde{V}%
(-p,\theta _{1},\overline{\theta }_{1}).\right\}  \notag
\end{align}%
The above expression contain quadratic, linear and logarithmic divergences.
However,we have cancellation of the quadratic and linear divergences coming
from other similar supergraphs. The overall divergences due to chiral
superfield give 
\begin{align}
& \overline{\Gamma }_{\Phi }(V) \\
& =\frac{1}{2}\sum_{A}T_{A}(R)g^{2}\int d^{4}\theta _{1}\frac{d^{4}p}{(2\pi
)^{4}}\frac{d^{4}k}{(2\pi )^{4}}  \notag \\
& \times \left\{ \frac{1}{2}\widetilde{V}^{a}(-p)\frac{%
p^{2}P_{T}+2((k-p)^{2}+m_{n}^{2})-2(k-p)^{2}-2m_{n}^{2}}{%
[k^{2}+m_{n_{i}}^{2}][(k-p)^{2}+m_{n_{i}}^{2}]}\widetilde{V}^{a}(p)\right\} 
\notag
\end{align}%
where $P_{T}=\frac{D^{\beta }\overline{D}^{2}D_{\beta }}{8p^{2}}.$ We are
then left with only logarithmic divergence.%
\begin{align}
\overline{\Gamma }_{\Phi }(V)& =\frac{1}{2}\sum_{A}T_{A}(R)g^{2}\int \frac{%
d^{4}p}{(2\pi )^{4}}d^{4}\theta _{1}\left[ \frac{1}{2}\widetilde{V}%
^{a}(-p)p^{2}P_{T}\widetilde{V}^{a}(p)\right]  \notag \\
& \times \left\{ \int \frac{d^{d}k\mu ^{4-d}}{(2\pi )^{d}}\frac{1}{%
[k^{2}+m_{n_{i}}^{2}][(k-p)^{2}+m_{n_{i}}^{2}]}\right\} .
\end{align}%
With the aid of dimensional reduction and method used by reference \cite%
{DDG-intscale}, the loop integration can be evaluated in d-dimension. The
correction reduces to 
\begin{align}
\overline{\Gamma }_{\Phi }(V)& =\frac{g^{2}}{2(4\pi )^{2}}%
\sum_{A}T(R)\int_{0}^{1}dz\int \frac{d^{4}p}{(2\pi )^{4}} \\
& \times \int_{0}^{\infty }\frac{dt}{t}e^{-t[pz(1-z)+m_{0}^{2}]}\int
d^{4}\theta \left\{ \frac{1}{2}\widetilde{V}^{a}(-p)p^{2}P_{T}\widetilde{V}%
^{a}(p)\right\} .  \notag
\end{align}%
The vector superfield can propagate in the extra dimensions, hence will have
KK contribution. Their contribution to the effective action is figure 2,
takes on the form%
\begin{align}
\overline{\Gamma }_{V,KK}(V)& =-\frac{g^{2}}{2}\frac{5}{2}%
\sum_{n_{i}=-\infty }^{\infty }C_{2}(g)\int \frac{d^{4}p}{(2\pi )^{4}}\int
d^{4}\theta \left\{ \frac{1}{2}\widetilde{V}^{a}(-p)p^{2}P_{T}\widetilde{V}%
^{a}(p)\right\}  \notag \\
& \times \left\{ \int \frac{d^{d}k\mu ^{4-d}}{(2\pi )^{d}}\frac{1}{%
[k^{2}+m_{n_{i}}^{2}][(k-p)^{2}+m_{n_{i}}^{2}]}\right\}
\end{align}%
where $\sum_{n_{i}=-\infty }^{\infty }=\sum_{n_{1}=-\infty }^{\infty
}\sum_{n_{2}=-\infty }^{\infty }\cdot \cdot \cdot \sum_{n_{\delta }=-\infty
}^{\infty },$ is the summation over KK states and $\delta $ is the number of
extra dimensions. $C_{2}(g)$ is the second Casimir coefficient. After
reparametrization of the denominator, the correction becomes 
\begin{align}
\overline{\Gamma }_{V,KK}(V)& =-\frac{g^{2}}{2}\frac{5}{2}%
\sum_{n_{i}=-\infty }^{\infty }C_{2}(g)\int_{0}^{1}dz\int \frac{d^{4}p}{%
(2\pi )^{4}}\int d^{4}\theta \left\{ \frac{1}{2}\widetilde{V}%
^{a}(-p)p^{2}P_{T}\widetilde{V}^{a}(p)\right\}  \notag \\
& \times \left\{ \int \frac{d^{d}k\mu ^{4-d}}{(2\pi )^{d}}\frac{1}{%
[k^{2}-2k\cdot pz+p^{2}z+m_{n_{i}}^{2}]^{2}}\right\}
\end{align}%
We can perform the loop integration in d-dimension . Using 
\begin{equation}
\int \frac{d^{d}p}{[p^{2}+2p\cdot q-M^{2}]^{\alpha }}=\frac{(-1)^{d}\pi ^{%
\frac{d}{2}}\Gamma (\alpha -\frac{d}{2})}{\Gamma (\alpha
)[-q^{2}-M^{2}]^{\alpha -\frac{d}{2}}}
\end{equation}%
and let $q=-pz$ and $-M^{2}=p^{2}z+m_{n_{i}}^{2}$ .The correction reduces to%
\begin{align}
\overline{\Gamma }_{V,KK}(V)& =-\frac{g^{2}}{2}\frac{5}{2}%
\sum_{n_{i}=-\infty }^{\infty }C_{2}(g)\int_{0}^{1}dz\int \frac{d^{4}p}{%
(2\pi )^{4}}\int d^{4}\theta \left\{ \frac{1}{2}\widetilde{V}%
^{a}(-p)p^{2}P_{T}\widetilde{V}^{a}(p)\right\}  \notag \\
& \times \left\{ \frac{(-1)^{\frac{d}{2}}\mu ^{4-d}}{(2\pi )^{d}}\frac{\pi ^{%
\frac{d}{2}}\Gamma (2-\frac{d}{2})}{\Gamma
(2)[-(pz)^{2}+p^{2}z+m_{n_{i}}^{2}]^{2-\frac{d}{2}}}\right\} ,
\end{align}%
letting $\varepsilon =4-d$ the correction reduces to%
\begin{align}
\overline{\Gamma }_{V,KK}(V)& =-\frac{g^{2}}{2}\frac{5}{2}%
\sum_{n_{i}=-\infty }^{\infty }C_{2}(g)\int_{0}^{1}dz\int \frac{d^{4}p}{%
(2\pi )^{4}}\int d^{4}\theta \left\{ \frac{1}{2}\widetilde{V}%
^{a}(-p)p^{2}P_{T}\widetilde{V}^{a}(p)\right\}  \notag \\
& \times \left\{ \frac{(-1)^{\frac{d}{2}}\mu ^{\varepsilon }}{(2\pi )^{d}}%
\frac{\pi ^{\frac{d}{2}}\Gamma (\frac{\varepsilon }{2})}{\Gamma
(2)[p^{2}z(1-z)+m_{n_{i}}^{2}]^{\frac{\varepsilon }{2}}}\right\} .
\end{align}%
By using the identity%
\begin{equation}
\int_{0}^{\infty }x^{n}e^{-ax}dx=\frac{\Gamma (n+1)}{a^{n+1}}
\end{equation}%
let $a=p^{2}z(1-z)+m_{n_{i}}^{2}$ and $\frac{\varepsilon }{2}=n+1$, the
correction becomes%
\begin{align}
\overline{\Gamma }_{V,KK}(V)& =-\frac{g^{2}}{2}\frac{5}{2}%
\sum_{n_{i}=-\infty }^{\infty }C_{2}(g)\int_{0}^{1}dz\int \frac{d^{4}p}{%
(2\pi )^{4}}\int d^{4}\theta \left\{ \frac{1}{2}\widetilde{V}%
^{a}(-p)p^{2}P_{T}\widetilde{V}^{a}(p)\right\}  \notag \\
& \times \left\{ \frac{(-1)^{\frac{d}{2}}\mu ^{\varepsilon }}{(4\pi )^{\frac{%
d}{2}}}\frac{\Gamma (\frac{\varepsilon }{2})}{\Gamma (\frac{\varepsilon }{2})%
}\int_{0}^{\infty }t^{\frac{\varepsilon }{2}-1}e^{-\left[
p^{2}z(1-z)+m_{n_{i}}^{2}\right] t}dt\right\} .
\end{align}%
Analytically continue from $d\rightarrow 4$ or $\varepsilon \rightarrow 0,$
the correction becomes%
\begin{align}
\overline{\Gamma }_{V,KK}(V)& =-\frac{g^{2}}{2}\frac{5}{2}%
\sum_{n_{i}=-\infty }^{\infty }C_{2}(g)\int_{0}^{1}dz\int \frac{d^{4}p}{%
(2\pi )^{4}}\int d^{4}\theta \left\{ \frac{1}{2}\widetilde{V}%
^{a}(-p)p^{2}P_{T}\widetilde{V}^{a}(p)\right\}  \notag \\
& \times \left\{ \frac{1}{(4\pi )^{2}}\int_{0}^{\infty }t^{-1}e^{-\left[
p^{2}z(1-z)+m_{n_{i}}^{2}\right] t}dt\right\} .
\end{align}%
Then we sum over the Kaluza-Klein states with the help of the Jacobi Theta
function 
\begin{align}
\overline{\Gamma }_{V,KK}(V)& =-\frac{g^{2}}{2}\frac{5}{2}\frac{1}{(4\pi
)^{2}}C_{2}(g)\int_{0}^{1}dz\int \frac{d^{4}p}{(2\pi )^{4}} \\
& \times \int_{0}^{\infty }\frac{dt}{t}\left\{ \Theta _{3}(\frac{it}{\pi
R^{2}})\right\} ^{\delta }e^{-p^{2}z(1-z)t}\int d^{4}\theta \left\{ \frac{1}{%
2}\widetilde{V}^{a}(-p)p^{2}P_{T}\widetilde{V}^{a}(p)\right\}  \notag
\end{align}%
where $\Theta _{3}$ is the Jacobi Theta function 
\begin{equation}
\Theta _{3}\left( \frac{it}{\pi R^{2}}\right) =\sum_{n_{i}=-\infty }^{\infty
}e^{(-m_{n_{i}}^{2}t)},
\end{equation}%
and $f_{ast}f_{bst}=\delta _{ab}C_{2}(g).$ Since the ghost superfields are
non-physical, it will not have ghost Kaluza-Klein contribution. The ghost
correction is figure 3%
\begin{align}
\overline{\Gamma }_{c}(V)& =-\frac{g^{2}}{2}\frac{1}{2}\frac{1}{(4\pi )^{2}}%
C_{2}(g)\int_{0}^{1}dz\int \frac{d^{4}p}{(2\pi )^{4}}\int_{0}^{\infty }\frac{%
dt}{t}e^{-[p^{2}z(1-z)+m_{0}^{2}]t}  \notag \\
& \times \int d^{4}\theta \left\{ \frac{1}{2}\widetilde{V}^{a}(-p)p^{2}P_{T}%
\widetilde{V}^{a}(p)\right\} .
\end{align}%
Therefore, the overall divergence at first order approximation can be
written as 
\begin{align}
\overline{\Gamma }_{KK,V,\Phi ,c}^{1loop}(V)& =\frac{g^{2}}{8\pi ^{2}}%
\int_{0}^{1}dz\int \frac{d^{4}p}{(2\pi )^{4}}\int_{0}^{\infty }\frac{dt}{t}%
e^{-p^{2}z(1-z)t}  \notag \\
& \times \left\{ 
\begin{array}{c}
e^{(-m_{0}^{2}t)}\sum T_{A}(R)-\frac{5}{2}C_{2}(g)\left\{ \Theta _{3}(\frac{%
it}{\pi R^{2}})\right\} ^{\delta } \\ 
-\frac{1}{2}e^{(-m_{0}^{2}t)}C_{2}(g)%
\end{array}%
\right\}  \notag \\
& \times \int d^{4}\theta \left\{ \frac{1}{2}\widetilde{V}^{a}(-p)p^{2}P_{T}%
\widetilde{V}^{a}(p)\right\} ,
\end{align}%
where $\frac{D^{\beta }\overline{D}^{2}D_{\beta }}{8}=p^{2}P_{T}.$ The first
term in the bracket is the contribution of the chiral superfields, the
second term comes from the vector superfield, and the third term is from the
unphysical ghost superfield.

We are now in position to obtain the counter term and the renormalization
constant for our model.%
\begin{equation}
\Gamma _{effective}=\int d^{8}p\widetilde{V}^{a}\frac{D^{\beta }\overline{D}%
^{2}D_{\beta }}{8}\widetilde{V}^{a}+\overline{\Gamma }_{KK,V,\Phi ,c}^{1loop}
\end{equation}%
which is infinite. We must add a counter term to the effective action to
render it finite. 
\begin{align}
\Gamma & =\int dp^{8}\widetilde{V}^{a}(-p)\frac{D^{\beta }\overline{D}%
^{2}D_{\beta }}{8}\widetilde{V}^{a}(p)+\overline{\Gamma }_{KK,V,\Phi
,c}^{1loop}(V)  \notag \\
& +\Delta Z_{2}\int dp^{8}\widetilde{V}^{a}\frac{D^{\beta }\overline{D}%
^{2}D_{\beta }}{8}\widetilde{V}^{a}  \notag \\
& =\int dp^{8}(1+\Delta Z_{2})\widetilde{V}^{a}(-p)\frac{D^{\beta }\overline{%
D}^{2}D_{\beta }}{8}\widetilde{V}^{a}(p)+\overline{\Gamma }_{KK,V,\Phi
,c}^{1loop}(V)
\end{align}%
by inspection,$\Delta Z_{2}$ must be 
\begin{align}
\Delta Z_{2}& =-\frac{g^{2}}{8\pi ^{2}}\int_{0}^{1}dze^{e^{-p^{2}z(1-z)t}}%
\int_{0}^{\infty }\frac{dt}{t}  \notag \\
& \times \left\{ 
\begin{array}{c}
e^{-m_{0}^{2}t}\sum T_{A}(R)-\frac{5}{2}C_{2}(g)\left\{ \Theta _{3}(\frac{it%
}{\pi R^{2}})\right\} ^{\delta } \\ 
-\frac{1}{2}e^{-m_{0}^{2}t}C_{2}(g)%
\end{array}%
\right\} .
\end{align}%
At zero momentum transferred $p=0,$ we have 
\begin{equation}
\Delta Z_{2}=-\frac{g^{2}}{8\pi ^{2}}\int_{0}^{\infty }\frac{dt}{t}\left\{ 
\begin{array}{c}
e^{-m_{0}^{2}t}\sum T_{A}(R)-\frac{5}{2}C_{2}(g)\left\{ \Theta _{3}(\frac{it%
}{\pi R^{2}})\right\} ^{\delta } \\ 
-\frac{1}{2}e^{-m_{0}^{2}t}C_{2}(g)%
\end{array}%
\right\}
\end{equation}%
let $Z_{2}=1+\Delta Z_{2},$ $g_{r}=Z_{g}g,$and $Z_{g}Z_{2}^{\frac{1}{2}}=1$.
This implies $Z_{g}=Z_{2}^{-\frac{1}{2}}$ 
\begin{equation*}
Z_{g}=\left[ 1-\frac{g^{2}}{8\pi ^{2}}\int_{0}^{\infty }\frac{dt}{t}\left\{ 
\begin{array}{c}
e^{-m_{0}^{2}t}\sum T_{A}(R) \\ 
-\frac{5}{2}C_{2}(g)\left\{ \Theta _{3}(\frac{it}{\pi R^{2}})\right\}
^{\delta }-\frac{1}{2}e^{-m_{0}^{2}t}C_{2}(g)%
\end{array}%
\right\} \right] ^{-\frac{1}{2}}
\end{equation*}%
we then have $g^{2}=Z_{2}g_{r}^{2}\Longrightarrow $%
\begin{align}
\frac{1}{g_{r}^{2}}& =\frac{1}{g^{2}}Z_{2}  \notag \\
& =\frac{1}{g^{2}}\left[ 
\begin{array}{c}
1-\frac{g^{2}}{8\pi ^{2}}\int_{0}^{\infty }\frac{dt}{t} \\ 
\times \left\{ 
\begin{array}{c}
e^{-m_{0}^{2}t}\sum T_{A}(R)-\frac{5}{2}C_{2}(g)\left\{ \Theta _{3}(\frac{it%
}{\pi R^{2}})\right\} ^{\delta } \\ 
-\frac{1}{2}e^{-m_{0}^{2}t}C_{2}(g)%
\end{array}%
\right\}%
\end{array}%
\right]
\end{align}%
with $\alpha \equiv \frac{g^{2}}{4\pi }$, we then have the following%
\begin{equation*}
\alpha _{r}^{-1}=\alpha ^{-1}-\frac{1}{2\pi }\int_{0}^{\infty }\frac{dt}{t}%
\left\{ 
\begin{array}{c}
e^{-m_{0}^{2}t}\sum T_{A}(R) \\ 
-\frac{5}{2}C_{2}(g)\left\{ \Theta _{3}(\frac{it}{\pi R^{2}})\right\}
^{\delta } \\ 
-\frac{1}{2}e^{-m_{0}^{2}t}C_{2}(g)%
\end{array}%
\right\} .
\end{equation*}%
To obtain an explicit expression for the beta function,operate $t\frac{%
\partial }{\partial t}$ on both sides of%
\begin{align}
\frac{1}{g^{2}}& =\frac{1}{g_{r}^{2}}+\frac{1}{8\pi ^{2}}\int_{0}^{\infty }%
\frac{dt}{t}  \notag \\
& \times \left\{ 
\begin{array}{c}
e^{-m_{0}^{2}t}\sum T_{A}(R)-\frac{5}{2}C_{2}(g)\left\{ \Theta _{3}(\frac{it%
}{\pi R^{2}})\right\} ^{\delta } \\ 
-\frac{1}{2}e^{-m_{0}^{2}t}C_{2}(g)%
\end{array}%
\right\} .
\end{align}%
we have%
\begin{equation}
\frac{-2}{g^{3}}t\frac{\partial g}{\partial t}=\frac{1}{8\pi ^{2}}\int d%
\left[ \left\{ 
\begin{array}{c}
e^{-m_{0}^{2}t}\sum T_{A}(R)-\frac{5}{2}C_{2}(g)\left\{ \Theta _{3}(\frac{it%
}{\pi R^{2}})\right\} ^{\delta } \\ 
-\frac{1}{2}e^{-m_{0}^{2}t}C_{2}(g)%
\end{array}%
\right\} \right] .
\end{equation}%
The beta function becomes%
\begin{equation}
\beta (g)=t\frac{\partial g}{\partial t}=-\frac{g^{3}}{16\pi ^{2}}\left\{ 
\begin{array}{c}
e^{-m_{0}^{2}t}\sum T_{A}(R)-\frac{5}{2}C_{2}(g)\left\{ \Theta _{3}(\frac{it%
}{\pi R^{2}})\right\} ^{\delta } \\ 
-\frac{1}{2}e^{-m_{0}^{2}t}C_{2}(g)%
\end{array}%
\right\} .
\end{equation}%
In order to obtain results which agree with ref.\cite{DDG-intscale}, we must
modify our model by allowing the chiral or matter superfields to propagate
freely in the extra-dimensions. The correction to the effective action at
one-loop becomes 
\begin{align}
\overline{\Gamma }_{KK,V,\Phi ,c}^{1loop}(V)& =\frac{g^{2}}{8\pi ^{2}}%
\int_{0}^{1}dz\int \frac{d^{4}p}{(2\pi )^{4}}\int_{0}^{\infty }\frac{dt}{t}%
e^{-p^{2}z(1-z)t}  \notag \\
& \times \left\{ 
\begin{array}{c}
\left[ \sum T_{A}(R)-\frac{5}{2}C_{2}(g)\right] \left\{ \Theta _{3}(\frac{it%
}{\pi R^{2}})\right\} ^{\delta } \\ 
-\frac{1}{2}e^{(-m_{0}^{2}t)}C_{2}(g)%
\end{array}%
\right\} .  \notag \\
& \times \int d^{4}\theta \left\{ \frac{1}{2}\widetilde{V}^{a}(-p)p^{2}P_{T}%
\widetilde{V}^{a}(p)\right\} ,
\end{align}%
The renormalization constant is 
\begin{equation}
Z_{g}=\left[ 1-\frac{g^{2}}{8\pi ^{2}}\int_{0}^{\infty }\frac{dt}{t}\left\{ 
\begin{array}{c}
\left[ \sum T_{A}(R)-\frac{5}{2}C_{2}(g)\right] \left\{ \Theta _{3}(\frac{it%
}{\pi R^{2}})\right\} ^{\delta } \\ 
-\frac{1}{2}e^{-m_{0}^{2}t}C_{2}(g)%
\end{array}%
\right\} \right] .
\end{equation}%
The beta function becomes%
\begin{equation}
\beta (g)=t\frac{\partial g}{\partial t}=-\frac{g^{3}}{16\pi ^{2}}\left\{ 
\begin{array}{c}
\left[ \sum T_{A}(R)-\frac{5}{2}C_{2}(g)\right] \left\{ \Theta _{3}(\frac{it%
}{\pi R^{2}})\right\} ^{\delta } \\ 
-\frac{1}{2}e^{-m_{0}^{2}t}C_{2}(g)%
\end{array}%
\right\} .
\end{equation}%
The explicit agreement of the results to ref.\cite{DDG-intscale} is
transparent when we expand the effective action into component form\footnote{%
Inverse Fourier transforming $\int dp^{8}\widetilde{V}^{a}(-p)\frac{D^{\beta
}\overline{D}^{2}D_{\beta }}{8}\widetilde{V}^{a}(p)$ to configuration space
and expanding into component fields by the machinery of supersymmetry
algebra.}.

\section{Kaluza-Klein Contribution at Two-Loop}

In this section, we use supergraph technique to calculate three
super-diagrams contributing to the effective action. These diagrams contain
the massive chiral-multiplet correction to the massive vector-multiplet at
two-loop with Kaluza-Klein states. In the next section, we evaluate the
two-loop integral.

\subsection{Supergraph calculation}

In momentum space, the correction of figure 4 is%
\begin{align}
\overline{\Gamma }_{KK,1}^{2loop}(V)& =\sum_{n_{i}=-\infty }^{\infty
}g^{4}\int d^{8}z_{1}d^{8}z_{2}d^{8}z_{3}d^{8}z_{4}V(z_{1},\theta _{1},%
\overline{\theta }_{1})V(z_{4},\theta _{4},\overline{\theta }_{4}) \\
& \times \left\{ 
\begin{array}{c}
\lbrack D^{2}(z_{1})\frac{\delta ^{8}(z_{1}-z_{2})}{16(\partial
_{1}^{2}-m_{n_{i}}^{2})}\overleftarrow{\overline{D}^{2}}(z_{2})][D^{2}(z_{2})%
\frac{\delta ^{8}(z_{2}-z_{3})}{16(\partial _{2}^{2}-m_{n_{i}}^{2})}%
\overleftarrow{\overline{D}^{2}}(z_{3})] \\ 
\times \lbrack \frac{-\delta ^{8}(z_{3}-z_{2})}{16(\partial
_{3}^{2}-m_{n_{i}}^{2})}][D^{2}(z_{3})\frac{\delta ^{8}(z_{3}-z_{4})}{%
16(\partial _{3}^{2}-m_{n_{i}}^{2})}\overleftarrow{\overline{D}^{2}}(z_{4})]
\\ 
\times \lbrack D^{2}(z_{4})\frac{\delta ^{8}(z_{4}-z_{1})}{16(\partial
_{4}^{2}-m_{n_{i}}^{2})}\overleftarrow{\overline{D}^{2}}(z_{1})]%
\end{array}%
\right\}   \notag
\end{align}%
using $D_{1}^{2}\delta ^{8}(z_{1}-z_{2})\overleftarrow{\overline{D}^{2}}%
_{2}=D_{1}^{2}\overline{D}_{1}^{2}\delta ^{8}(z_{1}-z_{2}),$ the correction
yields

\begin{align}
\overline{\Gamma }_{KK,1}^{2loop}(V)& =\sum_{n_{i}=-\infty }^{\infty
}g^{4}\int d^{8}z_{1}d^{8}z_{2}d^{8}z_{3}d^{8}z_{4}V(z_{1},\theta _{1},%
\overline{\theta }_{1})V(z_{4},\theta _{4},\overline{\theta }_{4}) \\
& \times \left\{ 
\begin{array}{c}
\lbrack D^{2}(z_{1})\overline{D}^{2}(z_{1})\frac{\delta ^{8}(z_{1}-z_{2})}{%
16(\partial _{1}^{2}-m_{n_{i}}^{2})}][D^{2}(z_{2})\overline{D}^{2}(z_{2})%
\frac{\delta ^{8}(z_{2}-z_{3})}{16(\partial _{2}^{2}-m_{n_{i}}^{2})}] \\ 
\times \lbrack \frac{-\delta ^{8}(z_{3}-z_{2})}{16(\partial
_{3}^{2}-m_{n_{i}}^{2})}][D^{2}(z_{3})\overline{D}^{2}(z_{3})\frac{\delta
^{8}(z_{3}-z_{4})}{16(\partial _{3}^{2}-m_{n_{i}}^{2})}] \\ 
\times \lbrack D^{2}(z_{4})\overline{D}^{2}(z_{4})\frac{\delta
^{8}(z_{4}-z_{1})}{16(\partial _{4}^{2}-m_{n_{i}}^{2})}]%
\end{array}%
\right\} .  \notag
\end{align}%
Using the result at one-loop, 
\begin{equation}
D_{1}^{2}\overline{D}_{1}^{2}\delta _{12}^{8}D_{2}^{2}\overline{D}%
_{2}^{2}\delta _{23}^{8}=\delta _{12}^{8}\left[ 16^{2}\square \delta
_{23}^{4}+i\frac{16^{2}}{2}\partial _{\overset{\cdot }{\sigma }}^{\alpha
}\delta _{23}^{4}\overline{D}_{2}^{\overset{\cdot }{\sigma }}D_{\alpha
2}+16\delta _{23}^{4}\overline{D}_{2}^{2}D_{2}^{2}\right] 
\end{equation}%
where $\delta _{12}^{8}=\delta ^{8}(z_{1}-z_{2})=\delta
^{4}(x_{1}-x_{2})\delta ^{4}(\theta _{1}-\theta _{2}),$ the correction
reduces to 
\begin{align}
\overline{\Gamma }_{KK,1}^{2loop}(V)& =\sum_{n_{i}=-\infty }^{\infty
}-g^{4}\int [\Pi _{j=1}^{4}d^{4}x_{j}d^{4}\theta _{j}]V(z_{1},\theta _{1},%
\overline{\theta }_{1})V(z_{4},\theta _{4},\overline{\theta }_{4}) \\
& \times \left\{ 
\begin{array}{c}
\delta ^{4}(x_{1}-x_{2})\delta ^{4}(\theta _{1}-\theta _{2})[16^{2}\square
\delta ^{4}(x_{2}-x_{3}) \\ 
+i\frac{16^{2}}{2}\partial _{\overset{\cdot }{\sigma }}^{\alpha }\delta
^{4}(x_{2}-x_{3})\overline{D}^{\overset{\cdot }{\sigma }}D_{\alpha
}+16\delta ^{4}(x_{2}-x_{3})\overline{D}^{2}D^{2}]%
\end{array}%
\right\}   \notag \\
& \times \left\{ 
\begin{array}{c}
\delta ^{8}(z_{3}-z_{2})\delta ^{8}(z_{3}-z_{4})[16^{2}\square \delta
^{4}(x_{4}-x_{1}) \\ 
+i\frac{16^{2}}{2}\partial _{\overset{\cdot }{\beta }}^{\beta }\delta
^{4}(x_{4}-x_{1})\overline{D}^{\overset{\cdot }{\beta }}D_{\beta }+16\delta
^{4}(x_{4}-x_{1})\overline{D}^{2}D^{2}]%
\end{array}%
\right\}   \notag \\
& \times \frac{16^{-5}}{(\square _{1}-m_{n_{i}}^{2})(\square
_{2}-m_{n_{i}}^{2})(\square _{3}-m_{n_{i}}^{2})(\square
_{4}-m_{n_{i}}^{2})(\square _{5}-m_{n_{i}}^{2})},  \notag
\end{align}%
\begin{align}
\overline{\Gamma }_{KK,1}^{2loop}(V)& =\sum_{n_{i}=-\infty }^{\infty
}-Sg^{4}\int [\Pi _{j=1}^{4}d^{4}x_{j}d^{4}\theta _{j}]V(z_{1},\theta _{1},%
\overline{\theta }_{1})V(z_{4},\theta _{4},\overline{\theta }_{4})  \notag \\
& \times \delta ^{4}(x_{1}-x_{2})\delta ^{4}(\theta _{1}-\theta _{2})\delta
^{4}(x_{2}-x_{3})  \notag \\
& \times \left\{ 
\begin{array}{c}
\lbrack 16^{2}(-q^{2})+\frac{16^{2}}{2}q_{\overset{\cdot }{\sigma }}^{\alpha
}\overline{D}^{\overset{\cdot }{\sigma }}D_{\alpha }+16\overline{D}^{2}D^{2}]
\\ 
\times \delta ^{8}(z_{3}-z_{2})\delta ^{8}(z_{3}-z_{4})\delta
^{4}(x_{4}-x_{1}) \\ 
\times \lbrack 16^{2}(-k^{2})+\frac{16^{2}}{2}k_{\overset{\cdot }{\beta }%
}^{\beta }\overline{D}^{\overset{\cdot }{\beta }}D_{\beta }+16\overline{D}%
^{2}D^{2}]%
\end{array}%
\right\}  \\
& \times \frac{16^{-5}}{(\square _{1}-m_{n_{i}}^{2})(\square
_{2}-m_{n_{i}}^{2})(\square _{3}-m_{n_{i}}^{2})(\square
_{4}-m_{n_{i}}^{2})(\square _{5}-m_{n_{i}}^{2})}.  \notag
\end{align}%
The bracket above yields%
\begin{align*}
& \delta _{12}^{8}\delta _{23}^{4}\{\} \\
& =\delta _{12}^{8}\delta _{23}^{4}\left\{ 
\begin{array}{c}
16^{4}q^{2}k^{2}\delta _{32}^{8}\delta _{34}^{8}\delta _{41}^{4}-\frac{16^{4}%
}{2}q^{2}k_{\overset{\cdot }{\beta }}^{\beta }\delta _{32}^{8}\delta
_{34}^{8}\delta _{41}^{4} \\ 
-16^{3}q^{2}\delta _{32}^{8}\delta _{34}^{8}\delta _{41}^{4}\overline{D}%
^{2}D^{2}+\frac{16^{4}}{2}q_{\overset{\cdot }{\sigma }}^{\alpha }\overline{D}%
^{\overset{\cdot }{\sigma }}D_{\alpha }\delta _{32}^{8}\delta
_{34}^{8}\delta _{41}^{4} \\ 
+\frac{16^{4}}{4}q_{\overset{\cdot }{\sigma }}^{\alpha }k_{\overset{\cdot }{%
\beta }}^{\beta }\overline{D}^{\overset{\cdot }{\sigma }}D_{\alpha }\delta
_{32}^{8}\delta _{34}^{8}\delta _{41}^{4}\overline{D}^{\overset{\cdot }{%
\beta }}D_{\beta }+\frac{16^{3}}{2}q_{\overset{\cdot }{\sigma }}^{\alpha }%
\overline{D}^{\overset{\cdot }{\sigma }}D_{\alpha }\delta _{32}^{8}\delta
_{34}^{8}\delta _{41}^{4}\overline{D}^{2}D^{2} \\ 
-16^{3}k^{2}\overline{D}^{2}D^{2}\delta _{32}^{8}\delta _{34}^{8}\delta
_{41}^{4}+\frac{16^{3}}{2}k_{\overset{\cdot }{\beta }}^{\beta }\overline{D}%
^{2}D^{2}\delta _{32}^{8}\delta _{34}^{8}\delta _{41}^{4}\overline{D}^{%
\overset{\cdot }{\beta }}D_{\beta } \\ 
+16^{2}\overline{D}^{2}D^{2}\delta _{32}^{8}\delta _{34}^{8}\delta _{41}^{4}%
\overline{D}^{2}D^{2}%
\end{array}%
\right\} 
\end{align*}%
In order to have non-vanishing result, we must have an even number of $%
D^{^{\prime }}s$ and $\overline{D}^{^{\prime }}s$ between the delta
functions. The fourth, fifth,and sixth term in the bracket vanish as
consequence of the identity. Next, we Fourier transform the vector-multiplet
superfield. With integration by parts and various properties of the
covariant derivatives and $\delta $-functions, the 2-loop correction reduces
to%
\begin{align}
\overline{\Gamma }_{KK,1}^{2loop}& =\sum_{n_{i}=-\infty }^{\infty
}-g^{4}\int [\Pi _{j=1}^{4}d^{4}x_{j}d^{4}\theta _{j}]\frac{d^{4}p}{(2\pi
)^{4}}\frac{d^{4}p^{\prime }}{(2\pi )^{4}}\widetilde{V}(-p,\theta _{1},%
\overline{\theta }_{1})\widetilde{V}(p^{\prime },\theta _{4},\overline{%
\theta }_{4})  \notag \\
& e^{-ip\cdot x_{1}}e^{ip^{\prime }\cdot x_{4}}  \notag \\
& \times \left\{ 
\begin{array}{c}
\delta _{12}^{8}\delta _{23}^{4}\delta _{32}^{8}\delta _{34}^{8}\delta
_{41}^{4}\left( 16^{4}(q\cdot k)^{2}-\frac{16^{4}}{2}iq^{2}k_{\overset{\cdot 
}{\beta }}^{\beta }\overline{D}^{\overset{\cdot }{\beta }}D_{\beta
}-16^{3}q^{2}\overline{D}^{2}D^{2}\right)  \\ 
+\delta _{12}^{8}\delta _{23}^{4}\delta _{32}^{4}\delta _{34}^{8}\delta
_{41}^{4}\left( -16^{4}k^{2}+\frac{16^{4}i}{2}r^{2}k_{\overset{\cdot }{\beta 
}}^{\beta }\overline{D}^{\overset{\cdot }{\beta }}D_{\beta }-16^{3}\overline{%
D}^{2}D^{2}\right) 
\end{array}%
\right\}   \notag \\
& \times \frac{16^{-5}}{%
(h^{2}+m_{n_{i}}^{2})(q^{2}+m_{n_{i}}^{2})(r^{2}+m_{n_{i}}^{2})(t^{2}+m_{n_{i}}^{2})(k^{2}+m_{n_{i}}^{2})%
}
\end{align}%
by inspection the second term in $\{\}$ contains only two $\delta
^{4}(\theta _{i}-\theta _{j})$ , hence it vanishes upon $\int d^{4}\theta
_{2}d^{4}\theta _{3}d^{4}\theta _{4}.$ The correction reduces to a single
point in superspace (i.e. $\theta _{1}$). Writing the explicit form of the
delta functions, the correction becomes%
\begin{align}
\overline{\Gamma }_{KK,1}^{2loop}(V)& =\sum_{n_{i}=-\infty }^{\infty
}-g^{4}\int d^{4}\theta _{1}d^{4}x_{1}d^{4}x_{2}d^{4}x_{3}d^{4}x_{4}%
\widetilde{V}(-p,\theta _{1},\overline{\theta }_{1})\widetilde{V}(p^{\prime
},\theta _{1},\overline{\theta }_{1})  \notag \\
& \times \int \frac{d^{4}h}{(2\pi )^{4}}e^{ih(x_{1}-x_{2})}\int \frac{d^{4}q%
}{(2\pi )^{4}}e^{iq(x_{2}-x_{3})}\int \frac{d^{4}r}{(2\pi )^{4}}%
e^{ir(x_{3}-x_{2})}  \notag \\
& \times \int \frac{d^{4}t}{(2\pi )^{4}}e^{it(x_{3}-x_{4})}\int \frac{d^{4}k%
}{(2\pi )^{4}}e^{ik(x_{4}-x_{1})} \\
& \times \left( 16^{4}(q\cdot k)^{2}-\frac{16^{4}}{2}iq^{2}k_{\overset{\cdot 
}{\beta }}^{\beta }\overline{D}^{\overset{\cdot }{\beta }}D_{\beta
}-16^{3}q^{2}\overline{D}^{2}D^{2}\right)   \notag \\
& \times \frac{16^{-5}}{%
(h^{2}+m_{n_{i}}^{2})(q^{2}+m_{n_{i}}^{2})(r^{2}+m_{n_{i}}^{2})(t^{2}+m_{n_{i}}^{2})(k^{2}+m_{n_{i}}^{2})%
}.  \notag
\end{align}%
All of the exponential terms can be rewritten so that we can integrate in
configuration space%
\begin{align}
\overline{\Gamma }_{KK,1}^{2loop}(V)& =\sum_{n_{i}=-\infty }^{\infty
}-g^{4}\int d^{4}\theta _{1}d^{4}x_{1}d^{4}x_{2}d^{4}x_{3}d^{4}x_{4}%
\widetilde{V}(-p,\theta _{1},\overline{\theta }_{1})\widetilde{V}(p^{\prime
},\theta _{1},\overline{\theta }_{1})  \notag \\
& \times \frac{d^{4}p}{(2\pi )^{4}}\frac{d^{4}p^{\prime }}{(2\pi )^{4}}\frac{%
d^{4}h}{(2\pi )^{4}}\frac{d^{4}q}{(2\pi )^{4}}\frac{d^{4}r}{(2\pi )^{4}}%
\frac{d^{4}t}{(2\pi )^{4}}\frac{d^{4}k}{(2\pi )^{4}}  \notag \\
& \times
e^{ix_{1}(h-p-k)}e^{ix_{2}(q-h-r)}e^{ix_{3}(r+t-q)}e^{ix_{4}(p^{\prime
}+k-t)} \\
& \times \left( 16^{4}(q\cdot k)^{2}-\frac{16^{4}}{2}iq^{2}k_{\overset{\cdot 
}{\beta }}^{\beta }\overline{D}^{\overset{\cdot }{\beta }}D_{\beta
}-16^{3}q^{2}\overline{D}^{2}D^{2}\right)   \notag \\
& \times \frac{16^{-5}}{%
(h^{2}+m_{n_{i}}^{2})(q^{2}+m_{n_{i}}^{2})(r^{2}+m_{n_{i}}^{2})(t^{2}+m_{n_{i}}^{2})(k^{2}+m_{n_{i}}^{2})%
},  \notag
\end{align}%
so we can perform $\int d^{4}x_{1}d^{4}x_{2}d^{4}x_{3}d^{4}x_{4}.$ This will
yield four delta functions%
\begin{align}
\overline{\Gamma }_{KK,1}^{2loop}(V)& =\sum_{n_{i}=-\infty }^{\infty
}-g^{4}\int d^{4}\theta _{1}\widetilde{V}(-p,\theta _{1},\overline{\theta }%
_{1})\widetilde{V}(p^{\prime },\theta _{1},\overline{\theta }_{1})  \notag \\
& \times \frac{d^{4}p}{(2\pi )^{4}}\frac{d^{4}p^{\prime }}{(2\pi )^{4}}\frac{%
d^{4}h}{(2\pi )^{4}}d^{4}qd^{4}rd^{4}td^{4}k \\
& \times \delta ^{4}(h-p-k)\delta ^{4}(q-h-r)\delta ^{4}(r+p-q)\delta
^{4}(p^{\prime }+k-t)  \notag \\
& \times \left\{ \frac{16^{4}(q\cdot k)^{2}-\frac{16^{4}}{2}iq^{2}k_{\overset%
{\cdot }{\beta }}^{\beta }\overline{D}^{\overset{\cdot }{\beta }}D_{\beta
}-16^{3}q^{2}\overline{D}^{2}D^{2}}{%
16^{5}(h^{2}+m_{n_{i}}^{2})(q^{2}+m_{n_{i}}^{2})(r^{2}+m_{n_{i}}^{2})(t^{2}+m_{n_{i}}^{2})(k^{2}+m_{n_{i}}^{2})%
}\right\} .  \notag
\end{align}%
Integrating over the delta functions one at a time, the correction yields%
\begin{align}
& \overline{\Gamma }_{KK,1}^{2loop}(V) \\
& =\sum_{n_{i}=-\infty }^{\infty }-g^{4}\int d^{4}\theta _{1}\widetilde{V}%
(-p,\theta _{1},\overline{\theta }_{1})\widetilde{V}(p,\theta _{1},\overline{%
\theta }_{1})\frac{d^{4}p}{(2\pi )^{4}}\frac{d^{4}r}{(2\pi )^{4}}\frac{d^{4}k%
}{(2\pi )^{4}}  \notag \\
& \times \left\{ \frac{\left[ 
\begin{array}{c}
16^{4}(r+p+k)^{2}k^{2}-\frac{16^{4}}{2}i(r+p+k)^{2}k_{\overset{\cdot }{\beta 
}}^{\beta }\overline{D}^{\overset{\cdot }{\beta }}D_{\beta } \\ 
-16^{3}(r+p+k)^{2}\overline{D}^{2}D^{2}%
\end{array}%
\right] }{16^{5}\left[ 
\begin{array}{c}
((p+k)^{2}+m_{n_{i}}^{2})((r+p+k)^{2}+m_{n_{i}}^{2})(r^{2}+m_{n_{i}}^{2}) \\ 
\times ((p+k)^{2}+m_{n_{i}}^{2})(k^{2}+m_{n_{i}}^{2})%
\end{array}%
\right] }\right\} .  \notag
\end{align}%
The above expression is the contribution of figure 4, to the effective
action. The contribution of other diagrams can be calculated the same way.

The contribution of figure 5 is%
\begin{align}
\overline{\Gamma }_{KK,2}^{2loop}(V)& =\sum_{n_{i}=-\infty }^{\infty
}g^{4}\int d^{8}z_{1}d^{8}z_{2}d^{8}z_{3}V(z_{1},\theta _{1},\overline{%
\theta }_{1})V(z_{3},\theta _{3},\overline{\theta }_{3}) \\
& \left\{ 
\begin{array}{c}
\lbrack D^{2}(z_{1})\frac{\delta ^{8}(z_{1}-z_{2})}{16(\partial
_{1}^{2}-m_{n_{i}}^{2})}\overleftarrow{\overline{D}^{2}}(z_{2})][\frac{%
-\delta ^{8}(z_{2}-z_{1})}{16(\partial _{2}^{2}-m_{n_{i}}^{2})}] \\ 
\lbrack D^{2}(z_{2})\frac{\delta ^{8}(z_{2}-z_{3})}{16(\partial
_{2}^{2}-m_{n_{i}}^{2})}\overleftarrow{\overline{D}^{2}}(z_{3})][D^{2}(z_{3})%
\frac{\delta ^{8}(z_{3}-z_{1})}{16(\partial _{2}^{2}-m_{n_{i}}^{2})}%
\overleftarrow{\overline{D}^{2}}(z_{1})]%
\end{array}%
\right\}   \notag
\end{align}%
\begin{align}
\overline{\Gamma }_{KK,2}^{2loop}(V)& =\sum_{n_{i}=-\infty }^{\infty
}g^{4}\int d^{8}z_{1}d^{8}z_{2}d^{8}z_{3}V(z_{1},\theta _{1},\overline{%
\theta }_{1})V(z_{3},\theta _{3},\overline{\theta }_{3}) \\
& \times \left\{ 
\begin{array}{c}
\lbrack D^{2}(z_{1})\overline{D}^{2}(z_{1})\frac{\delta ^{8}(z_{1}-z_{2})}{%
16(\partial _{1}^{2}-m_{n_{i}}^{2})}][\frac{-\delta ^{8}(z_{2}-z_{1})}{%
16(\partial _{2}^{2}-m_{n_{i}}^{2})}] \\ 
\lbrack D^{2}(z_{2})\overline{D}^{2}(z_{2})\frac{\delta ^{8}(z_{2}-z_{3})}{%
16(\partial _{2}^{2}-m_{n_{i}}^{2})}][D^{2}(z_{3})\overline{D}^{2}(z_{3})%
\frac{\delta ^{8}(z_{3}-z_{1})}{16(\partial _{2}^{2}-m_{n_{i}}^{2})}]%
\end{array}%
\right\}   \notag
\end{align}%
again using result at one loop, the correction reduces to%
\begin{align}
\overline{\Gamma }_{KK,2}^{2loop}(V)& =\sum_{n_{i}=-\infty }^{\infty
}g^{4}\int d^{8}z_{1}d^{8}z_{2}d^{8}z_{3}V(z_{1},\theta _{1},\overline{%
\theta }_{1})V(z_{3},\theta _{3},\overline{\theta }_{3})  \notag \\
& \times \left\{ 
\begin{array}{c}
\lbrack \frac{D_{1}^{2}\overline{D}_{1}^{2}\delta _{12}^{8}}{16(\partial
_{1}^{2}-m_{n_{i}}^{2})}][\frac{-\delta _{21}^{8}}{16(\partial
_{2}^{2}-m_{n_{i}}^{2})}] \\ 
\times \frac{\delta _{23}^{8}[16^{2}\square \delta _{31}^{4}+i\frac{16^{2}}{2%
}\partial _{\overset{\cdot }{\sigma }}^{\alpha }\delta _{31}^{4}\overline{D}%
^{\overset{\cdot }{\sigma }}D_{\alpha }+16\delta _{31}^{4}\overline{D}%
^{2}D^{2}]}{16^{2}(\partial _{3}^{2}-m_{n_{i}}^{2})(\partial
_{4}^{2}-m_{n_{i}}^{2})}%
\end{array}%
\right\} .
\end{align}%
With the aid of identity $(D_{1}^{2}\overline{D}_{1}^{2}\delta
_{12}^{8})\delta _{21}^{8}\delta _{23}^{8}=\delta _{12}^{8}\overline{D}%
_{1}^{2}D_{1}^{2}\delta _{21}^{8}\delta _{23}^{8},$ the correction reduces to%
\begin{align}
\overline{\Gamma }_{KK,2}^{2loop}(V)& =\sum_{n_{i}=-\infty }^{\infty
}g^{4}\int d^{8}z_{1}d^{8}z_{2}d^{8}z_{3}V(z_{1},\theta _{1},\overline{%
\theta }_{1})V(z_{3},\theta _{3},\overline{\theta }_{3})\delta _{12}^{8}%
\overline{D}_{1}^{2}D_{1}^{2}\delta _{21}^{8}\delta _{23}^{8}  \notag \\
& \times \left\{ \frac{\lbrack 16^{2}\square \delta _{31}^{4}+i\frac{16^{2}}{%
2}\partial _{\overset{\cdot }{\sigma }}^{\alpha }\delta _{31}^{4}\overline{D}%
^{\overset{\cdot }{\sigma }}D_{\alpha }+16\delta _{31}^{4}\overline{D}%
^{2}D^{2}]}{16^{4}[(\partial _{1}^{2}-m_{n_{i}}^{2})(\partial
_{2}^{2}-m_{n_{i}}^{2})(\partial _{3}^{2}-m_{n_{i}}^{2})(\partial
_{4}^{2}-m_{n_{i}}^{2})]}\right\} 
\end{align}%
\begin{align}
& \overline{\Gamma }_{KK,2}^{2loop}(V) \\
& =\sum_{n_{i}=-\infty }^{\infty }g^{4}\int
d^{8}z_{1}d^{8}z_{2}d^{8}z_{3}V(z_{1},\theta _{1},\overline{\theta }%
_{1})V(z_{3},\theta _{3},\overline{\theta }_{3})\delta _{12}^{8}\delta
_{21}^{4}\delta _{23}^{8}  \notag \\
& \times \left\{ \frac{\lbrack 16^{2}\square \delta _{31}^{4}+i\frac{16^{2}}{%
2}\partial _{\overset{\cdot }{\sigma }}^{\alpha }\delta _{31}^{4}\overline{D}%
^{\overset{\cdot }{\sigma }}D_{\alpha }+16\delta _{31}^{4}\overline{D}%
^{2}D^{2}]}{16^{3}[(\partial _{1}^{2}-m_{n_{i}}^{2})(\partial
_{2}^{2}-m_{n_{i}}^{2})(\partial _{3}^{2}-m_{n_{i}}^{2})(\partial
_{4}^{2}-m_{n_{i}}^{2})]}\right\} ,  \notag
\end{align}%
where we use $\delta _{12}^{8}\overline{D}_{1}^{2}D_{1}^{2}\delta
_{21}^{8}=16\delta _{12}^{8}\delta _{21}^{4}.$ Next we Fourier transform to
momentum space%
\begin{align}
\overline{\Gamma }_{KK,2}^{2loop}(V)& =\sum_{n_{i}=-\infty }^{\infty
}g^{4}\int d^{8}z_{1}d^{8}z_{2}d^{8}z_{3}\widetilde{V}(-p,\theta _{1},%
\overline{\theta }_{1})\widetilde{V}(p^{\prime },\theta _{3},\overline{%
\theta }_{3})  \notag \\
& \times \frac{d^{4}p}{(2\pi )^{4}}\frac{d^{4}p^{\prime }}{(2\pi )^{4}}\frac{%
d^{4}h}{(2\pi )^{4}}\frac{d^{4}q}{(2\pi )^{4}}\frac{d^{4}r}{(2\pi )^{4}}%
\frac{d^{4}k}{(2\pi )^{4}} \\
& \times e^{ix_{1}(h-r-k-p)}e^{ix_{2}(r-h+q)}e^{ix_{3}(p^{\prime
}+k-q)}\delta ^{4}(\theta _{1}-\theta _{2})\delta ^{4}(\theta _{2}-\theta
_{3})  \notag \\
& \times \left\{ \frac{\lbrack \frac{-1}{16}k^{2}+\frac{i}{32}\partial _{%
\overset{\cdot }{\sigma }}^{\alpha }\overline{D}^{\overset{\cdot }{\sigma }%
}D_{\alpha }+\frac{1}{16^{2}}\overline{D}^{2}D^{2}]}{%
[(h^{2}+m_{n_{i}}^{2})(r^{2}+m_{n_{i}}^{2})(q^{2}+m_{n_{i}}^{2})(k^{2}+m_{n_{i}}^{2})]%
}\right\} .  \notag
\end{align}%
Performing $\int \int d^{4}\theta _{2}d^{4}\theta _{3},$ the correction
yields 
\begin{align}
\overline{\Gamma }_{KK,2}^{2loop}(V)& =\sum_{n_{i}=-\infty }^{\infty
}g^{4}\int d^{8}z_{1}d^{4}x_{2}d^{4}x_{3}\widetilde{V}(-p,\theta _{1},%
\overline{\theta }_{1})\widetilde{V}(p^{\prime },\theta _{1},\overline{%
\theta }_{1})  \notag \\
& \times \frac{d^{4}p}{(2\pi )^{4}}\frac{d^{4}p^{\prime }}{(2\pi )^{4}}\frac{%
d^{4}h}{(2\pi )^{4}}\frac{d^{4}q}{(2\pi )^{4}}\frac{d^{4}r}{(2\pi )^{4}}%
\frac{d^{4}k}{(2\pi )^{4}}  \notag \\
& \times e^{ix_{1}(h-r-k-p)}e^{ix_{2}(r-h+q)}e^{ix_{3}(p^{\prime }+k-q)} \\
& \times \left\{ \frac{\lbrack \frac{-1}{16}k^{2}+\frac{i}{32}\partial _{%
\overset{\cdot }{\sigma }}^{\alpha }\overline{D}^{\overset{\cdot }{\sigma }%
}D_{\alpha }+\frac{1}{16^{2}}\overline{D}^{2}D^{2}]}{%
[(h^{2}+m_{n_{i}}^{2})(r^{2}+m_{n_{i}}^{2})(q^{2}+m_{n_{i}}^{2})(k^{2}+m_{n_{i}}^{2})]%
}\right\} ,  \notag
\end{align}%
where the expression reduces to a single point in superspace. We then
systematically integrate over the coordinate space $\int
d^{4}x_{1}d^{4}x_{2}d^{4}x_{3}.$ This will give us three delta functions in
momentum space%
\begin{align}
\overline{\Gamma }_{KK,2}^{2loop}(V)& =\sum_{n_{i}=-\infty }^{\infty
}g^{4}\int d^{8}\theta _{1}\widetilde{V}(-p,\theta _{1},\overline{\theta }%
_{1})\widetilde{V}(p^{\prime },\theta _{1},\overline{\theta }_{1})  \notag \\
& \times \frac{d^{4}p}{(2\pi )^{4}}\frac{d^{4}p^{\prime }}{(2\pi )^{4}}\frac{%
d^{4}h}{(2\pi )^{4}}d^{4}rd^{4}qd^{4}k\delta ^{4}(h-r-k-p)  \notag \\
& \times \delta ^{4}(r-h+q)\delta ^{4}(p^{\prime }+k-q) \\
& \times \left\{ \frac{\lbrack \frac{-1}{16}k^{2}+\frac{i}{32}\partial _{%
\overset{\cdot }{\sigma }}^{\alpha }\overline{D}^{\overset{\cdot }{\sigma }%
}D_{\alpha }+\frac{1}{16^{2}}\overline{D}^{2}D^{2}]}{%
[(h^{2}+m_{n_{i}}^{2})(r^{2}+m_{n_{i}}^{2})(q^{2}+m_{n_{i}}^{2})(k^{2}+m_{n_{i}}^{2})]%
}\right\} .  \notag
\end{align}%
Integrating over the three delta functions $\int d^{4}hd^{4}qd^{4}p^{\prime
},$ The contribution of diagram yields \ 
\begin{align}
\overline{\Gamma }_{KK,2}^{2loop}(V)& =\sum_{n_{i}=-\infty }^{\infty
}-g^{4}\int d^{4}\theta _{1}\frac{d^{4}p}{(2\pi )^{4}}\frac{d^{4}r}{(2\pi
)^{4}}\frac{d^{4}k}{(2\pi )^{4}}\widetilde{V}(-p,\theta _{1},\overline{%
\theta }_{1})\widetilde{V}(p,\theta _{1},\overline{\theta }_{1}) \\
& \times \left\{ \frac{\lbrack \frac{-1}{16}k^{2}+\frac{i}{32}k_{\overset{%
\cdot }{\beta }}^{\beta }\overline{D}^{\overset{\cdot }{\beta }}D_{\beta }-%
\frac{1}{16^{2}}\overline{D}^{2}D^{2}]}{%
((r+p+k)^{2}+m_{n_{i}}^{2})(r^{2}+m_{n_{i}}^{2})((p+k)^{2}+m_{n_{i}}^{2})(k^{2}+m_{n_{i}}^{2})%
}\right\} .  \notag
\end{align}

The contribution of figure 6 is%
\begin{align}
\overline{\Gamma }_{KK,3}^{2loop}(V)& =\sum_{n_{i}=-\infty }^{\infty
}g^{4}\int d^{8}z_{1}d^{8}z_{2}V(z_{1},\theta _{1},\overline{\theta }%
_{1})V(z_{2},\theta _{2},\overline{\theta }_{2}) \\
& \left\{ [\frac{D_{1}^{2}\overline{D}_{1}^{2}\delta _{12}^{8}}{16(\partial
_{1}^{2}-m_{n_{i}}^{2})}][\frac{-\delta _{21}^{8}}{16(\partial
_{2}^{2}-m_{n_{i}}^{2})}][\frac{D_{2}^{2}\overline{D}_{2}^{2}\delta _{21}^{8}%
}{16(\partial _{1}^{2}-m_{n_{i}}^{2})}]\right\} .  \notag
\end{align}%
Using properties of super-spinor derivatives and integration by parts and
the following 
\begin{equation*}
(D_{1}^{2}\overline{D}_{1}^{2}\delta _{12}^{8})\delta _{21}^{8}(D_{2}^{2}%
\overline{D}_{2}^{2}\delta _{21}^{8})=16^{2}\delta _{12}^{8}\delta
_{12}^{4}\delta _{21}^{4},
\end{equation*}%
the correction becomes%
\begin{align}
\overline{\Gamma }_{KK,3}^{2loop}(V)& =-\sum_{n_{i}=-\infty }^{\infty
}g^{4}\int d^{8}z_{1}d^{8}z_{2}V(z_{1},\theta _{1},\overline{\theta }%
_{1})V(z_{2},\theta _{2},\overline{\theta }_{2})  \notag \\
& \times \left\{ \frac{\delta _{12}^{8}\delta _{12}^{4}\delta _{21}^{4}}{%
16(\partial _{1}^{2}-m_{n_{i}}^{2})(\partial
_{2}^{2}-m_{n_{i}}^{2})(\partial _{3}^{2}-m_{n_{i}}^{2})}\right\} .
\end{align}%
Fourier transforms into momentum space and integrate $\int d^{4}\theta _{2}$%
\begin{align}
\overline{\Gamma }_{KK,3}^{2loop}(V)& =-\sum_{n_{i}=-\infty }^{\infty
}g^{4}\int d^{4}\theta _{1}d^{4}x_{1}d^{4}x_{2}\widetilde{V}(-p,\theta _{1},%
\overline{\theta }_{1})\widetilde{V}(p^{\prime },\theta _{1},\overline{%
\theta }_{1})  \notag \\
& \times \frac{d^{4}p}{(2\pi )^{4}}\frac{d^{4}p^{\prime }}{(2\pi )^{4}}\frac{%
d^{4}h}{(2\pi )^{4}}\frac{d^{4}q}{(2\pi )^{4}}\frac{d^{4}r}{(2\pi )^{4}}%
e^{ix_{1}(h-p-r-q)} \\
& \times e^{ix_{2}(p^{\prime }+q-h+r)}\left\{ \frac{1}{%
16(h^{2}+m_{n_{i}}^{2})(r^{2}+m_{n_{i}}^{2})(q^{2}+m_{n_{i}}^{2})}\right\} .
\notag
\end{align}%
Performing $\int d^{4}x_{1}d^{4}x_{2}$%
\begin{align}
\overline{\Gamma }_{KK,3}^{2loop}(V)& =-\sum_{n_{i}=-\infty }^{\infty
}g^{4}\int d^{4}\theta _{1}\widetilde{V}(-p,\theta _{1},\overline{\theta }%
_{1})\widetilde{V}(p^{\prime },\theta _{1},\overline{\theta }_{1})  \notag \\
& \times \frac{d^{4}p}{(2\pi )^{4}}\frac{d^{4}p^{\prime }}{(2\pi )^{4}}\frac{%
d^{4}h}{(2\pi )^{4}}d^{4}rd^{4}q \\
& \times \left\{ \frac{\delta ^{4}(h-p-r-q)\delta ^{4}(p^{\prime }+q-h+r)}{%
16(h^{2}+m_{n_{i}}^{2})(r^{2}+m_{n_{i}}^{2})(q^{2}+m_{n_{i}}^{2})}\right\} ,
\notag
\end{align}%
and integrate $\int d^{4}hd^{4}p^{\prime }$ yields%
\begin{align}
\overline{\Gamma }_{KK,3}^{2loop}(V)& =-\sum_{n_{i}=-\infty }^{\infty
}g^{4}\int d^{4}\theta _{1}\frac{d^{4}p}{(2\pi )^{4}}\frac{d^{4}r}{(2\pi
)^{4}}\frac{d^{4}q}{(2\pi )^{4}}\widetilde{V}(-p,\theta _{1},\overline{%
\theta }_{1})\widetilde{V}(p,\theta _{1},\overline{\theta }_{1})  \notag \\
& \times \left\{ \frac{1}{%
16((p+q+r)^{2}+m_{n_{i}}^{2})(r^{2}+m_{n_{i}}^{2})(q^{2}+m_{n_{i}}^{2})}%
\right\} .
\end{align}

The contribution of figure 7 is%
\begin{align}
\overline{\Gamma }_{KK,4}^{2loop}(V)& =\sum_{n_{i}=-\infty }^{\infty
}g^{4}\int d^{8}z_{1}d^{8}z_{2}d^{8}z_{3}d^{8}z_{4}V(z_{1},\theta _{1},%
\overline{\theta }_{1})V(z_{3},\theta _{3},\overline{\theta }_{3})  \notag \\
& \times \left\{ 
\begin{array}{c}
\lbrack \frac{D_{1}^{2}\overline{D}_{1}^{2}\delta _{12}^{8}}{16(\partial
_{1}^{2}-m_{n_{i}}^{2})}][\frac{-\delta _{24}^{8}}{16(\partial
_{2}^{2}-m_{n_{i}}^{2})}][\frac{D_{2}^{2}\overline{D}_{2}^{2}\delta _{23}^{8}%
}{16(\partial _{3}^{2}-m_{n_{i}}^{2})}] \\ 
\times \lbrack \frac{D_{3}^{2}\overline{D}_{3}^{2}\delta _{34}^{8}}{%
16(\partial _{4}^{2}-m_{n_{i}}^{2})}][\frac{D_{4}^{2}\overline{D}%
_{4}^{2}\delta _{41}^{8}}{16(\partial _{5}^{2}-m_{n_{i}}^{2})}]%
\end{array}%
\right\} ,
\end{align}%
\begin{align}
\overline{\Gamma }_{KK,4}^{2loop}(V)& =-\sum_{n_{i}=-\infty }^{\infty
}g^{4}\int d^{8}z_{1}d^{8}z_{2}d^{8}z_{3}d^{8}z_{4}V(z_{1},\theta _{1},%
\overline{\theta }_{1})V(z_{3},\theta _{3},\overline{\theta }_{3}) \\
& \times (D_{1}^{2}\overline{D}_{1}^{2}\delta _{12}^{8})\delta
_{24}^{8}(D_{2}^{2}\overline{D}_{2}^{2}\delta _{23}^{8})\delta _{34}^{8} 
\notag \\
& \times \left\{ \frac{\lbrack 16^{2}\square \delta _{41}^{4}+i\frac{16^{2}}{%
2}\partial _{\overset{\cdot }{\sigma }}^{\alpha }\delta _{41}^{4}\overline{D}%
^{\overset{\cdot }{\sigma }}D_{\alpha }+16\delta _{41}^{4}\overline{D}%
^{2}D^{2}]}{16^{5}[(\partial _{1}^{2}-m_{n_{i}}^{2})(\partial
_{2}^{2}-m_{n_{i}}^{2})(\partial _{3}^{2}-m_{n_{i}}^{2})(\partial
_{4}^{2}-m_{n_{i}}^{2})(\partial _{5}^{2}-m_{n_{i}}^{2})]}\right\} .  \notag
\end{align}%
Using $\delta _{24}^{8}D_{2}^{2}\overline{D}_{2}^{2}\delta
_{23}^{8}=16\delta _{24}^{8}\delta _{23}^{4},$ the correction becomes%
\begin{align}
\overline{\Gamma }_{KK,4}^{2loop}(V)& =-\sum_{n_{i}=-\infty }^{\infty
}g^{4}\int d^{8}z_{1}d^{8}z_{2}d^{8}z_{3}d^{8}z_{4}V(z_{1},\theta _{1},%
\overline{\theta }_{1})V(z_{3},\theta _{3},\overline{\theta }_{3}) \\
& \times (D_{1}^{2}\overline{D}_{1}^{2}\delta _{12}^{8})\delta
_{24}^{8}\delta _{23}^{4}\delta _{34}^{8}  \notag \\
& \times \left\{ \frac{\lbrack 16^{2}\square \delta _{41}^{4}+i\frac{16^{2}}{%
2}\partial _{\overset{\cdot }{\sigma }}^{\alpha }\delta _{41}^{4}\overline{D}%
^{\overset{\cdot }{\sigma }}D_{\alpha }+16\delta _{41}^{4}\overline{D}%
^{2}D^{2}]}{16^{5}[(\partial _{1}^{2}-m_{n_{i}}^{2})(\partial
_{2}^{2}-m_{n_{i}}^{2})(\partial _{3}^{2}-m_{n_{i}}^{2})(\partial
_{4}^{2}-m_{n_{i}}^{2})(\partial _{5}^{2}-m_{n_{i}}^{2})]}\right\} .  \notag
\end{align}%
With the properties of the super-spinor derivatives and integration by
parts, and the following 
\begin{equation*}
(D_{1}^{2}\overline{D}_{1}^{2}\delta _{12}^{8})\delta _{24}^{8}\delta
_{23}^{4}\delta _{34}^{8}=32\delta _{12}^{8}\delta _{24}^{4}\delta
_{23}^{4}\delta _{34}^{8},
\end{equation*}%
the correction reduces to%
\begin{align}
\overline{\Gamma }_{KK,4}^{2loop}(V)& =-\sum_{n_{i}=-\infty }^{\infty
}g^{4}\int d^{8}z_{1}d^{8}z_{2}d^{8}z_{3}d^{8}z_{4}V(z_{1},\theta _{1},%
\overline{\theta }_{1})V(z_{3},\theta _{3},\overline{\theta }_{3}) \\
& \times 32\delta _{12}^{8}\delta _{24}^{4}\delta _{23}^{4}\delta _{34}^{8} 
\notag \\
& \left\{ \frac{[16^{2}\square \delta _{41}^{4}+i\frac{16^{2}}{2}\partial _{%
\overset{\cdot }{\sigma }}^{\alpha }\delta _{41}^{4}\overline{D}^{\overset{%
\cdot }{\sigma }}D_{\alpha }+16\delta _{41}^{4}\overline{D}^{2}D^{2}]}{%
16^{5}[(\partial _{1}^{2}-m_{n_{i}}^{2})(\partial
_{2}^{2}-m_{n_{i}}^{2})(\partial _{3}^{2}-m_{n_{i}}^{2})(\partial
_{4}^{2}-m_{n_{i}}^{2})(\partial _{5}^{2}-m_{n_{i}}^{2})]}\right\} .  \notag
\end{align}%
By inspection, the above expression contain only $\delta ^{4}(\theta
_{1}-\theta _{2})\delta ^{4}(\theta _{3}-\theta _{4}).$ Thus upon $\int
d^{4}\theta _{2}d^{4}\theta _{3}d^{4}\theta _{4}$ integration, the
correction vanishes%
\begin{equation}
\overline{\Gamma }_{KK,4}^{2loop}(V)=0
\end{equation}%
These are all possible diagrams of chiral correction to the vector
superfield.

\subsection{Evaluation of Two-Loop Integrals}

In this section, we compute the two-loop integrals of the non-vanishing
super-diagrams:

The evaluation of figure 4 begins with the result of the previous section
where superfield formalism was used to manipulate the contribution in to
manageable form. The correction is 
\begin{align}
& \Gamma _{KK,1}^{2loop}(V)  \notag \\
& =\sum_{n_{i}=-\infty }^{\infty }-g^{4}\int d^{4}\theta _{1}\widetilde{V}%
(-p,\theta _{1},\overline{\theta }_{1})\widetilde{V}(p,\theta _{1},\overline{%
\theta }_{1})\frac{d^{4}p}{(2\pi )^{4}}\frac{d^{4}r}{(2\pi )^{4}}\frac{d^{4}k%
}{(2\pi )^{4}}  \notag \\
& \times \left\{ \frac{\left[ 
\begin{array}{c}
16^{4}(r+p+k)^{2}k^{2}-\frac{16^{4}}{2}i(r+p+k)^{2}k_{\overset{\cdot }{\beta 
}}^{\beta }\overline{D}^{\overset{\cdot }{\beta }}D_{\beta } \\ 
-16^{3}(r+p+k)^{2}\overline{D}^{2}D^{2}%
\end{array}%
\right] }{16^{5}\left[ 
\begin{array}{c}
((p+k)^{2}+m_{n_{i}}^{2})((r+p+k)^{2}+m_{n_{i}}^{2})(r^{2}+m_{n_{i}}^{2}) \\ 
((p+k)^{2}+m_{n_{i}}^{2})(k^{2}+m_{n_{i}}^{2})%
\end{array}%
\right] }\right\} .
\end{align}%
We first expand the numerator and reparametrizing the denominator%
\begin{align}
& \overline{\Gamma }_{KK,1}^{2loop}(V) \\
& =\sum_{n_{i}=-\infty }^{\infty }-g^{4}\int d^{4}\theta _{1}\widetilde{V}%
(-p,\theta _{1},\overline{\theta }_{1})\widetilde{V}(p,\theta _{1},\overline{%
\theta }_{1})\frac{d^{4}p}{(2\pi )^{4}}\frac{d^{4}r}{(2\pi )^{4}}\frac{d^{4}k%
}{(2\pi )^{4}}N  \notag \\
& \times 4!\int_{0}^{1}dy_{1}dy_{2}dy_{3}dy_{4}\left[ 
\begin{array}{c}
y_{1}((p+k)^{2}+m_{n_{i}}^{2})+y_{2}((r+p+k)^{2}+m_{n_{i}}^{2}) \\ 
+y_{3}(r^{2}+m_{n_{i}}^{2})+y_{4}((p+k)^{2}+m_{n_{i}}^{2}) \\ 
+(1-y_{1}-y_{2}-y_{3}-y_{4})(k^{2}+m_{n_{i}}^{2})%
\end{array}%
\right] ^{-5},  \notag
\end{align}%
where $N$ is the expanded numerators. Using $\int_{0}^{\infty
}t^{n}e^{-at}dt=\frac{\Gamma (n+1)}{a^{n+1}},$ and completing the square in $%
r,$ the above becomes%
\begin{align}
& \Gamma _{KK,1}^{2loop}(V) \\
& =\sum_{n_{i}=-\infty }^{\infty }-g^{4}\int d^{4}\theta _{1}\widetilde{V}%
(-p,\theta _{1},\overline{\theta }_{1})\widetilde{V}(p,\theta _{1},\overline{%
\theta }_{1})\frac{d^{4}p}{(2\pi )^{4}}\frac{d^{4}r}{(2\pi )^{4}}\frac{d^{4}k%
}{(2\pi )^{4}}N4!\int_{0}^{\infty }t^{4}dt  \notag \\
& \frac{\int_{0}^{1}dy_{1}dy_{2}dy_{3}dy_{4}}{\Gamma (5)}e^{-\left\{ 
\begin{array}{c}
(y_{2}+y_{3})\left[ r+\frac{y_{2}(p+k)}{y_{2}+y_{3}}\right] ^{2}-\frac{%
y_{2}^{2}(p^{2}+k^{2}+2pk)}{y_{2}+y_{3}} \\ 
+k^{2}(1-y_{1}-y_{3}-y_{4})+2kp(y_{1}+y_{2}+y_{4})+ \\ 
p^{2}(y_{1}+y_{2}+y_{4})+m_{n}^{2}%
\end{array}%
\right\} t}.  \notag
\end{align}%
\ Shifting the integration variable to $r^{\prime }=r+\frac{y_{2}(p+k)}{%
y_{2}+y_{3}}\equiv r+U,$ the numerator becomes 
\begin{equation}
N=\frac{1}{16^{2}\cdot 2}\left[ 
\begin{array}{c}
r^{\prime 2}(32k^{2}-ik_{\overset{\cdot }{\beta }}^{\beta }\overline{D}^{%
\overset{\cdot }{\beta }}D_{\beta }-32\overline{D}^{2}D^{2}) \\ 
+32(U^{2}k^{2}+p^{2}k^{2}+k^{4}-2pUk^{2}+2pk^{3}-2Uk^{3}) \\ 
+32(-U^{2}-p^{2}-k^{2}+2pU-2pk+2Uk)\overline{D}^{2}D^{2} \\ 
-i(U^{2}+p^{2}+k^{2}-2pU+2pk-2Uk)k_{\overset{\cdot }{\beta }}^{\beta }%
\overline{D}^{\overset{\cdot }{\beta }}D_{\beta }%
\end{array}%
\right] .
\end{equation}%
The integration w.r.t. $r^{\prime }$ is carried out using $\int \frac{%
dr^{\prime d}}{(2\pi )^{d}}e^{-ar^{\prime 2}}=(4\pi a)^{-\frac{d}{2}}$ and $%
\int \frac{dr^{\prime d}}{(2\pi )^{d}}r^{\prime 2}e^{-ar^{\prime 2}}=\frac{%
(4\pi a)^{-\frac{d}{2}}}{2a}.$%
\begin{align}
& \overline{\Gamma }_{KK,1}^{2loop}(V) \\
& =\sum_{n_{i}=-\infty }^{\infty }-Sg^{4}\int d^{4}\theta _{1}\widetilde{V}%
(-p,\theta _{1},\overline{\theta }_{1})\widetilde{V}(p,\theta _{1},\overline{%
\theta }_{1})\frac{d^{4}p}{(2\pi )^{4}}\frac{d^{4}k}{(2\pi )^{4}}N^{\prime
}4!\int_{0}^{1}dy_{1}dy_{2}dy_{3}dy_{4}  \notag \\
& \times \frac{1}{\Gamma (5)}\int_{0}^{1}t^{4}dte^{-\left\{ -\frac{\left(
p(y_{2}+y_{4})+ky_{2}\right) ^{2}}{y_{2}+y_{3}+y_{4}}%
+k^{2}(1-y_{1}-y_{3}-y_{4})+2kp(y_{1}+y_{2})+p^{2}(y_{1}+y_{2}+y_{4})+m_{n}^{2}\right\} t}
\notag
\end{align}%
where%
\begin{align}
16^{2}\cdot 2N^{\prime }& =\frac{(32k^{2}-ik_{\overset{\cdot }{\beta }%
}^{\beta }\overline{D}^{\overset{\cdot }{\beta }}D_{\beta }-32\overline{D}%
^{2}D^{2})}{32\pi ^{2}(y_{2}+y_{3})^{3}t^{3}}+\frac{1}{16\pi
^{2}(y_{2}+y_{3})^{2}t^{2}}  \notag \\
& \times \left[ 
\begin{array}{c}
32(U^{2}k^{2}+p^{2}k^{2}+k^{4}-2pUk^{2}+2pk^{3}-2Uk^{3}) \\ 
+32(-U^{2}-p^{2}-k^{2}+2pU-2pk+2Uk)\overline{D}^{2}D^{2} \\ 
-i(U^{2}+p^{2}+k^{2}-2pU+2pk-2Uk)k_{\overset{\cdot }{\beta }}^{\beta }%
\overline{D}^{\overset{\cdot }{\beta }}D_{\beta }%
\end{array}%
\right] .
\end{align}%
Completing the square in $k$ and then shift integration variable to 
\begin{equation}
k^{\prime }=k+\frac{P\left( y_{1}+y_{2}+y_{4}-\frac{y_{2}^{2}}{y_{2}+y_{3}}%
\right) }{\left( 1-y_{1}-y_{3}-y_{4}-\frac{y_{2}^{2}}{y_{2}+y_{3}}\right) }%
\equiv k+M
\end{equation}%
The correction becomes%
\begin{align}
& \overline{\Gamma }_{KK,1}^{2loop}(V)  \notag \\
& =\sum_{n_{i}=-\infty }^{\infty }-g^{4}\int d^{4}\theta _{1}\widetilde{V}%
(-p,\theta _{1},\overline{\theta }_{1})\widetilde{V}(p,\theta _{1},\overline{%
\theta }_{1})\frac{d^{4}p}{(2\pi )^{4}}N^{\prime
}4!\int_{0}^{1}dy_{1}dy_{2}dy_{3}dy_{4}  \notag \\
& \frac{1}{\Gamma (5)}\int_{0}^{1}t^{4}dt\int_{0}^{\infty }\frac{%
d^{4}k^{\prime }}{(2\pi )^{4}} \\
& \times \exp \left\{ 
\begin{array}{c}
k^{\prime 2}\left( 1-y_{1}-y_{3}-y_{4}-\frac{y_{2}^{2}}{y_{2}+y_{3}}\right)
+m_{n}^{2} \\ 
+p^{2}\left( \frac{\left( y_{1}+y_{2}+y_{4}-\frac{y_{2}^{2}}{y_{2}+y_{3}}%
\right) ^{2}}{\left( 1-y_{1}-y_{3}-y_{4}-\frac{y_{2}^{2}}{y_{2}+y_{3}}%
\right) }+\left( y_{1}+y_{2}+y_{3}-\frac{y_{2}^{2}}{y_{2}+y_{3}}\right)
\right) 
\end{array}%
\right\} (-t).  \notag
\end{align}%
Integration w.r.t. $k^{\prime }$ yields%
\begin{align}
& \overline{\Gamma }_{KK,1}^{2loop}(V)  \notag \\
& =\sum_{n_{i}=-\infty }^{\infty }-g^{4}\int d^{4}\theta _{1}\widetilde{V}%
(-p,\theta _{1},\overline{\theta }_{1})\widetilde{V}(p,\theta _{1},\overline{%
\theta }_{1})\frac{d^{4}p}{(2\pi )^{4}}N^{\prime \prime }4!  \notag \\
& \int_{0}^{1}\frac{dy_{1}dy_{2}dy_{3}dy_{4}}{\Gamma (5)16^{2}\cdot 2\pi
^{2}(y_{2}+y_{3})^{2}}\int_{0}^{\infty }t^{2}dt  \notag \\
& \times e^{-\left[ p^{2}\left( \frac{\left( y_{1}+y_{2}+y_{4}-\frac{%
y_{2}^{2}}{y_{2}+y_{3}}\right) ^{2}}{\left( 1-y_{1}-y_{3}-y_{4}-\frac{%
y_{2}^{2}}{y_{2}+y_{3}}\right) }+\left( y_{1}+y_{2}+y_{3}-\frac{y_{2}^{2}}{%
y_{2}+y_{3}}\right) \right) +m_{n}^{2}\right] t},
\end{align}%
where 
\begin{align}
N^{\prime \prime }& =\frac{3}{64\pi ^{2}Y^{4}t^{4}}\left[ 2+\frac{2y_{2}^{2}%
}{(y_{2}+y_{3})^{2}}\right]  \\
& +\frac{1}{32\pi ^{2}Y^{3}t^{3}}\left[ 
\begin{array}{c}
\frac{1}{(y_{2}+y_{3})t}+2R^{2}+\frac{2M^{2}y_{2}^{2}}{(y_{2}+y_{3})^{2}}%
+12M^{2}+2M^{4}-4pR \\ 
+\frac{4pMy_{2}}{(y_{2}+y_{3})}-12pM+12RM-\frac{12M^{2}y_{2}}{(y_{2}+y_{3})}%
-2\overline{D}^{2}D^{2} \\ 
+i\left( \frac{3M}{16}-\frac{p}{8}+\frac{R}{8}-\frac{My_{2}}{8(y_{2}+y_{3})}%
\right) \overline{D}^{\overset{\cdot }{\beta }}D_{\beta } \\ 
+\frac{y_{2}^{2}}{(y_{2}+y_{3})^{2}}-\frac{2y_{2}^{2}\overline{D}^{2}D^{2}}{%
(y_{2}+y_{3})^{2}}%
\end{array}%
\right]   \notag \\
& +\frac{1}{16\pi ^{2}Y^{2}t^{2}}\left[ 
\begin{array}{c}
\frac{M^{2}}{(y_{2}+y_{3})t}+i\frac{M_{\overset{\cdot }{\beta }}^{\beta }%
\overline{D}^{\overset{\cdot }{\beta }}D_{\beta }}{32(y_{2}+y_{3})t}-\frac{%
\overline{D}^{2}D^{2}}{(y_{2}+y_{3})t} \\ 
+2M\left( R^{2}-\frac{2RMy_{2}}{(y_{2}+y_{3})}+\frac{M^{2}y_{2}^{2}}{%
(y_{2}+y_{3})^{2}}\right) -2pM+2M^{4} \\ 
-4pRM^{2}+\frac{4pM^{3}y_{2}}{(y_{2}+y_{3})}-4pM^{3}-4RM^{3}+\frac{%
4M^{4}y_{2}}{(y_{2}+y_{3})} \\ 
+(-2R^{2}+\frac{4RMy_{2}}{(y_{2}+y_{3})}-\frac{2M^{2}y_{2}^{2}}{%
(y_{2}+y_{3})^{2}}-2p^{2}-2M^{2}+4pR \\ 
-\frac{4pMy_{2}}{(y_{2}+y_{3})}+4pM-4RM+\frac{4M^{2}y_{2}}{(y_{2}+y_{3})})%
\overline{D}^{2}D^{2} \\ 
+i(\frac{RM_{\overset{\cdot }{\beta }}^{\beta }}{16}-\frac{y_{2}MM_{\overset{%
\cdot }{\beta }}^{\beta }}{16(y_{2}+y_{3})}+\frac{MM_{\overset{\cdot }{\beta 
}}^{\beta }}{16}+\frac{pRM_{\overset{\cdot }{\beta }}^{\beta }}{8} \\ 
-\frac{py_{2}MM_{\overset{\cdot }{\beta }}^{\beta }}{8(y_{2}+y_{3})}+\frac{%
pMM_{\overset{\cdot }{\beta }}^{\beta }}{8}-\frac{RMM_{\overset{\cdot }{%
\beta }}^{\beta }}{8}+\frac{y_{2}M^{2}M_{\overset{\cdot }{\beta }}^{\beta }}{%
8(y_{2}+y_{3})})\overline{D}^{\overset{\cdot }{\beta }}D_{\beta }%
\end{array}%
\right] .  \notag
\end{align}%
Here 
\begin{equation}
Y=\left( 1-y_{1}-y_{3}-y_{4}-\frac{y_{2}^{2}}{y_{2}+y_{3}}\right) ,
\end{equation}%
\begin{equation}
M=\frac{p\left( y_{1}+y_{2}+y_{4}-\frac{y_{2}^{2}}{y_{2}+y_{3}}\right) }{%
\left( 1-y_{1}-y_{3}-y_{4}-\frac{y_{2}^{2}}{y_{2}+y_{3}}\right) },
\end{equation}%
\begin{equation}
U=\frac{(p+k)y_{2}}{y_{2}+y_{3}}\equiv R+\frac{ky_{2}}{y_{2}+y_{3}},
\end{equation}%
\begin{equation}
R=\frac{py_{2}}{y_{2}+y_{3}}.
\end{equation}

Correction of figure 5 is%
\begin{align}
\overline{\Gamma }_{KK,2}^{2loop}(V)& =\sum_{n_{i}=-\infty }^{\infty
}-g^{4}\int d^{4}\theta _{1}\frac{d^{4}p}{(2\pi )^{4}}\frac{d^{4}r}{(2\pi
)^{4}}\frac{d^{4}k}{(2\pi )^{4}}\widetilde{V}(-p,\theta _{1},\overline{%
\theta }_{1})\widetilde{V}(p,\theta _{1},\overline{\theta }_{1}) \\
& \times \left\{ \frac{\left[ \frac{-1}{16}k^{2}+\frac{i}{32}k_{\overset{%
\cdot }{\beta }}^{\beta }\overline{D}^{\overset{\cdot }{\beta }}D_{\beta }-%
\frac{1}{16^{2}}\overline{D}^{2}D^{2}\right] }{%
((r+p+k)^{2}+m_{n_{i}}^{2})(r^{2}+m_{n_{i}}^{2})((p+k)^{2}+m_{n_{i}}^{2})(k^{2}+m_{n_{i}}^{2})%
}\right\} .  \notag
\end{align}%
Similarly, we reparametrized the denominator and cast in terms of an
exponential function. The correction yields%
\begin{align}
\Gamma _{KK,2}^{2loop}(V)& =\sum_{n_{i}=-\infty }^{\infty }-g^{4}\int
d^{4}\theta _{1}\frac{d^{4}p}{(2\pi )^{4}}\frac{d^{4}r^{\prime }}{(2\pi )^{4}%
}\frac{d^{4}k}{(2\pi )^{4}}\widetilde{V}(-p,\theta _{1},\overline{\theta }%
_{1})\widetilde{V}(p,\theta _{1},\overline{\theta }_{1})  \notag \\
& \times \frac{3!}{\Gamma (4)}\int_{0}^{1}\Pi
_{i=1}^{3}dy_{i}N\int_{0}^{\infty }dtt^{3} \\
& \times \exp \left\{ -\left[ 
\begin{array}{c}
(y_{1}+y_{2})r^{\prime 2}-\frac{y_{1}^{2}(p+k)^{2}}{y_{1}+y_{2}}%
+(1-y_{2})k^{2}+(y_{1}+y_{3})p^{2} \\ 
+2kp(y_{1}+y_{3})+2ry_{1}(p+k)+m_{n}^{2}%
\end{array}%
\right] t\right\} ,  \notag
\end{align}%
where $N=\left[ \frac{-1}{16}k^{2}+\frac{i}{32}k_{\overset{\cdot }{\beta }%
}^{\beta }\overline{D}^{\overset{\cdot }{\beta }}D_{\beta }-\frac{1}{16^{2}}%
\overline{D}^{2}D^{2}\right] .$ Integration w.r.t. $r^{\prime }$ yields%
\begin{align}
\overline{\Gamma }_{KK,2}^{2loop}(V)& =\sum_{n_{i}=-\infty }^{\infty
}-g^{4}\int d^{4}\theta _{1}\frac{d^{4}p}{(2\pi )^{4}}\frac{d^{4}k}{(2\pi
)^{4}}\widetilde{V}(-p,\theta _{1},\overline{\theta }_{1})\widetilde{V}%
(p,\theta _{1},\overline{\theta }_{1})  \notag \\
& \times \frac{3!}{\Gamma (4)}\int_{0}^{1}\frac{\Pi _{i=1}^{3}dy_{i}N}{16\pi
^{2}(y_{1}+y_{2})^{2}}\int_{0}^{\infty }dtt \\
& \times \exp \left\{ -\left[ 
\begin{array}{c}
(1-y_{2})k^{2}-\frac{y_{1}^{2}(p+k)^{2}}{y_{1}+y_{2}}+(y_{1}+y_{3})p^{2} \\ 
+2kp(y_{1}+y_{3})+2ry_{1}(p+k)+m_{n}^{2}%
\end{array}%
\right] t\right\} .  \notag
\end{align}%
Completing the square in $k$ and shift to $k^{\prime }=k+\left( \frac{%
y_{1}y_{2}+y_{3}y_{1}+y_{3}y_{2}}{y_{1}+y_{2}-y_{2}y_{1}-y_{2}^{2}-y_{1}^{2}}%
\right) p\equiv k+T.$ Integrating w.r.t $k^{\prime }$ yields%
\begin{align}
\overline{\Gamma }_{KK,2}^{2loop}(V)& =\sum_{n_{i}=-\infty }^{\infty
}-g^{4}\int d^{4}\theta _{1}\frac{d^{4}p}{(2\pi )^{4}}\widetilde{V}%
(-p,\theta _{1},\overline{\theta }_{1})\widetilde{V}(p,\theta _{1},\overline{%
\theta }_{1})\frac{3!}{\Gamma (4)16\pi ^{2}} \\
& \times \int_{0}^{\infty }dtt\exp \left\{ -\left[ p^{2}\left( 
\begin{array}{c}
y_{1}+y_{2}+\frac{y_{1}^{2}}{y_{1}+y_{2}} \\ 
+\frac{\left( y_{1}y_{2}+y_{3}y_{1}+y_{3}y_{2}\right) ^{2}}{%
(y_{1}+y_{2})\left( y_{1}+y_{2}-y_{2}y_{1}-y_{2}^{2}-y_{1}^{2}\right) }%
\end{array}%
\right) +m_{n}^{2}\right] t\right\}   \notag \\
& \times \left\{ 
\begin{array}{c}
\int_{0}^{1}\frac{\Pi _{i=1}^{3}dy_{i}(y_{1}+y_{2})}{16^{2}2\pi
^{2}t^{3}\left( y_{1}+y_{2}-y_{2}y_{1}-y_{2}^{2}-y_{1}^{2}\right) ^{3}} \\ 
+\int_{0}^{1}\frac{\Pi _{i=1}^{3}dy_{i}}{16^{2}\pi ^{2}t^{2}\left(
y_{1}+y_{2}-y_{2}y_{1}-y_{2}^{2}-y_{1}^{2}\right) }\left[ -T^{2}-\frac{iT_{%
\overset{\cdot }{\sigma }}^{\alpha }\overline{D}^{\overset{\cdot }{\sigma }%
}D_{\alpha }}{2}+\frac{\overline{D}^{2}D^{2}}{16}\right] 
\end{array}%
\right\} .  \notag
\end{align}%
The contribution from figure 6 is%
\begin{align}
\overline{\Gamma }_{KK,3}^{2loop}(V)& =-\sum_{n_{i}=-\infty }^{\infty
}g^{4}\int d^{4}\theta _{1}\frac{d^{4}p}{(2\pi )^{4}}\frac{d^{4}r}{(2\pi
)^{4}}\frac{d^{4}q}{(2\pi )^{4}}\widetilde{V}(-p,\theta _{1},\overline{%
\theta }_{1})\widetilde{V}(p,\theta _{1},\overline{\theta }_{1})  \notag \\
& \times \left\{ \frac{1}{%
16((p+q+r)^{2}+m_{n_{i}}^{2})(r^{2}+m_{n_{i}}^{2})(q^{2}+m_{n_{i}}^{2})}%
\right\} .
\end{align}%
The evaluation of this diagram yields%
\begin{align}
\overline{\Gamma }_{KK,3}^{2loop}(V)& =-\sum_{n_{i}=-\infty }^{\infty }\frac{%
g^{4}}{16^{3}\pi ^{4}\Gamma (3)}\int d^{4}\theta _{1}\frac{d^{4}p}{(2\pi
)^{4}}\widetilde{V}(-p,\theta _{1},\overline{\theta }_{1})\widetilde{V}%
(p,\theta _{1},\overline{\theta }_{1})  \notag \\
& \times \int_{0}^{1}\frac{dy_{1}dy_{2}}{(1-y_{2})^{2}\left( y_{1}+y_{2}-%
\frac{y_{1}^{2}}{1-y_{2}}\right) ^{2}}\int_{0}^{\infty }\frac{dt}{t^{2}} \\
& \times \exp \left\{ -\left[ p^{2}\left( y_{1}-\frac{y_{1}^{2}}{1-y_{2}}-%
\frac{\left( y_{1}-\frac{y_{1}^{2}}{1-y_{2}}\right) }{\left( y_{1}+y_{2}-%
\frac{y_{1}^{2}}{1-y_{2}}\right) }\right) +m_{n}^{2}\right] t\right\} . 
\notag
\end{align}%
Integration w.r.t. Feynman parameters $\{y_{i}\}$ are carried out using
maple. The correction to the effective action yields%
\begin{equation}
\overline{\Gamma }_{total}^{2loop}=\overline{\Gamma }_{1}^{2loop}+\overline{%
\Gamma }_{2}^{2loop}+\overline{\Gamma }_{3}^{2loop}.
\end{equation}%
Similar to the one-loop correction, we define $Z_{2}^{2loop}=1+\Delta
Z_{2}^{2loop}.$ At zero momentum transferred $p=0$, we calculated $%
Z_{2}^{2loop}$ to be%
\begin{align}
Z_{2}^{2loop}& =1+\sum_{n_{i}=-\infty }^{\infty }\frac{g^{4}4!}{16^{3}\Gamma
(5)2\pi ^{4}}  \notag \\
& \times \left\{ 
\begin{array}{c}
\frac{3}{2}\int_{0}^{\infty }\frac{dt}{t^{2}}e^{-m_{n}^{2}t}\left[ \frac{-27%
}{70}-\frac{1}{\varepsilon ^{5}}\left( \frac{1}{21\varepsilon ^{2}}+\frac{1}{%
6\varepsilon }+\frac{6}{5}\right) \right]  \\ 
+\frac{1}{2}\int_{0}^{\infty }\frac{dt}{t}e^{-m_{n}^{2}t}\left[ 
\begin{array}{c}
\left( \frac{-3}{10}-\frac{1}{5\varepsilon ^{5}}\right) \frac{1}{t}-2%
\overline{D}^{2}D^{2}\left( \frac{-3}{20}-\frac{1}{\varepsilon ^{4}}\left( 
\frac{1}{10\varepsilon }-\frac{1}{4}\right) \right)  \\ 
+(1-2\overline{D}^{2}D^{2})\left( 
\begin{array}{c}
\frac{-7}{9}+4\ln 2-\frac{3}{2\varepsilon ^{5}}-\frac{8}{9\varepsilon ^{3}}-%
\frac{1}{6\varepsilon ^{2}} \\ 
+\frac{1}{3\varepsilon }+\frac{1}{3}\ln \varepsilon 
\end{array}%
\right) 
\end{array}%
\right]  \\ 
-\int_{0}^{\infty }\frac{dt}{t}e^{-m_{n}^{2}t}\overline{D}^{2}D^{2}\left[ 
\frac{-7}{9}+4\ln 2-\frac{3}{2\varepsilon ^{5}}-\frac{8}{9\varepsilon ^{3}}-%
\frac{1}{6\varepsilon ^{2}}+\frac{1}{3\varepsilon }+\frac{1}{3}\ln
\varepsilon \right] 
\end{array}%
\right\}   \notag \\
& -\sum_{n_{i}=-\infty }^{\infty }\frac{Sg^{4}3!}{16^{3}\Gamma (5)\pi ^{4}} 
\notag \\
& \times \left\{ 
\begin{array}{c}
\frac{1}{2}\int_{0}^{\infty }\frac{dt}{t^{2}}e^{-m_{n}^{2}t}\left[ 
\begin{array}{c}
\frac{9877}{1536}-\frac{3}{64\varepsilon ^{2}}+\frac{1645}{192\varepsilon }-%
\frac{1277713}{1079129}\ln \varepsilon  \\ 
+\int_{0}^{1}L(y_{1})dy_{1}%
\end{array}%
\right]  \\ 
+\int_{0}^{\infty }\frac{dt}{t}e^{-m_{n}^{2}t}\left( -T^{2}-\frac{iT_{%
\overset{\cdot }{\sigma }}^{\alpha }\overline{D}^{\overset{\cdot }{\sigma }%
}D_{\alpha }}{2}+\frac{\overline{D}^{2}D^{2}}{16}\right)  \\ 
\times \left( 
\begin{array}{c}
\frac{-1}{2}+\frac{23}{24}\arctan h\left( \frac{4}{28}\sqrt{14}\right) -%
\frac{23}{24}\arctan h\left( \frac{4}{28}\sqrt{14}\right)  \\ 
+\frac{1}{2\varepsilon }+3\ln \varepsilon -\int_{0}^{1}K(y_{1})dy_{1}%
\end{array}%
\right) 
\end{array}%
\right\}   \notag \\
& -\sum_{n_{i}=-\infty }^{\infty }\frac{g^{4}2!}{16^{3}\Gamma (3)\pi ^{4}}%
\int_{0}^{\infty }\frac{dt}{t^{2}}e^{-m_{n}^{2}t}\left\{ 
\begin{array}{c}
\frac{-1}{\varepsilon }-2\ln \varepsilon +1-\frac{9}{4}\ln 4 \\ 
+\int_{0}^{1}F(y_{1})dy_{1}%
\end{array}%
\right\} 
\end{align}%
Summing over the Kaluza Klein modes yields%
\begin{align}
Z_{2}^{2loop}& =1+\frac{g^{4}4!}{16^{3}\Gamma (5)2\pi ^{4}}  \notag \\
& \left\{ 
\begin{array}{c}
\frac{3}{2}\int_{0}^{\infty }\frac{dt}{t^{2}}\left\{ \Theta _{3}(\frac{it}{%
\pi R^{2}})\right\} ^{\delta }\left[ \frac{-27}{70}-\frac{1}{\varepsilon ^{5}%
}\left( \frac{1}{21\varepsilon ^{2}}+\frac{1}{6\varepsilon }+\frac{6}{5}%
\right) \right]  \\ 
+\frac{1}{2}\int_{0}^{\infty }\frac{dt}{t}\left\{ \Theta _{3}(\frac{it}{\pi
R^{2}})\right\} ^{\delta }\left[ 
\begin{array}{c}
\left( \frac{-3}{10}-\frac{1}{5\varepsilon ^{5}}\right) \frac{1}{t}-2%
\overline{D}^{2}D^{2}\left( \frac{-3}{20}-\frac{1}{\varepsilon ^{4}}\left( 
\frac{1}{10\varepsilon }-\frac{1}{4}\right) \right)  \\ 
+(1-2\overline{D}^{2}D^{2})\left( 
\begin{array}{c}
\frac{-7}{9}+4\ln 2-\frac{3}{2\varepsilon ^{5}}-\frac{8}{9\varepsilon ^{3}}-%
\frac{1}{6\varepsilon ^{2}} \\ 
+\frac{1}{3\varepsilon }+\frac{1}{3}\ln \varepsilon 
\end{array}%
\right) 
\end{array}%
\right]  \\ 
-\int_{0}^{\infty }\frac{dt}{t}\left\{ \Theta _{3}(\frac{it}{\pi R^{2}}%
)\right\} ^{\delta }\overline{D}^{2}D^{2}\left[ \frac{-7}{9}+4\ln 2-\frac{3}{%
2\varepsilon ^{5}}-\frac{8}{9\varepsilon ^{3}}-\frac{1}{6\varepsilon ^{2}}+%
\frac{1}{3\varepsilon }+\frac{1}{3}\ln \varepsilon \right] 
\end{array}%
\right\} .  \notag \\
& -\frac{g^{4}3!}{16^{3}\Gamma (5)\pi ^{4}}  \notag \\
& \times \left\{ 
\begin{array}{c}
\frac{1}{2}\int_{0}^{\infty }\frac{dt}{t^{2}}\left\{ \Theta _{3}(\frac{it}{%
\pi R^{2}})\right\} ^{\delta }\left[ 
\begin{array}{c}
\frac{9877}{1536}-\frac{3}{64\varepsilon ^{2}}+\frac{1645}{192\varepsilon }-%
\frac{1277713}{1079129}\ln \varepsilon  \\ 
+\int_{0}^{1}L(y_{1})dy_{1}%
\end{array}%
\right]  \\ 
+\int_{0}^{\infty }\frac{dt}{t}\left\{ \Theta _{3}(\frac{it}{\pi R^{2}}%
)\right\} ^{\delta }\left( -T^{2}-\frac{iT_{\overset{\cdot }{\sigma }%
}^{\alpha }\overline{D}^{\overset{\cdot }{\sigma }}D_{\alpha }}{2}+\frac{%
\overline{D}^{2}D^{2}}{16}\right)  \\ 
\times \left( 
\begin{array}{c}
\frac{-1}{2}+\frac{23}{24}\arctan h\left( \frac{4}{28}\sqrt{14}\right) -%
\frac{23}{24}\arctan h\left( \frac{4}{28}\sqrt{14}\right)  \\ 
+\frac{1}{2\varepsilon }+3\ln \varepsilon -\int_{0}^{1}K(y_{1})dy_{1}%
\end{array}%
\right) 
\end{array}%
\right\}   \notag \\
& -\frac{g^{4}2!}{16^{3}\Gamma (3)\pi ^{4}}\int_{0}^{\infty }\frac{dt}{t^{2}}%
\left\{ \Theta _{3}(\frac{it}{\pi R^{2}})\right\} ^{\delta }\left\{ 
\begin{array}{c}
\frac{-1}{\varepsilon }-2\ln \varepsilon +1-\frac{9}{4}\ln 4 \\ 
+\int_{0}^{1}F(y_{1})dy_{1}%
\end{array}%
\right\} .
\end{align}

\section{Summary}

In this paper, we have embedded Kaluza-Klein excitations directly into $%
N=1,D=4$ MSSM via superfield formulation, where many component fields
diagrams are calculated simultaneously. The effects of the extradimensions
are transparent through the computation of the one-loop correction to the
vector superfield. As application, we calculated the Beta functions and
evolution of the gauge couplings derived here agree with the component field
approach. In addition, we calculate the two-loop partial corrections to
vector superfield where pure virtual vector loops corrections are not
considered.

\section{Appendix}

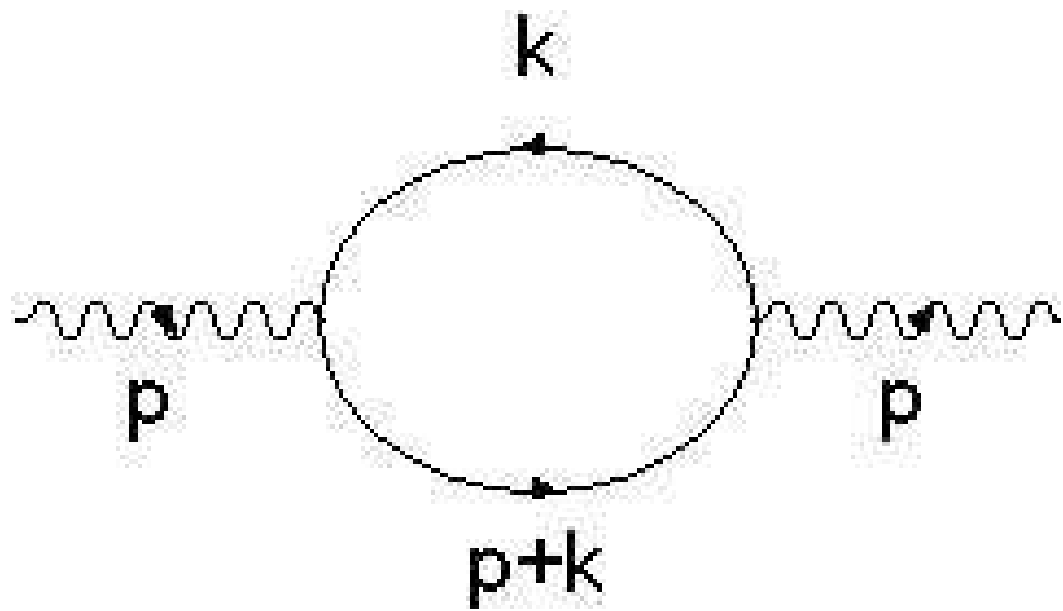
Figure 1: Chiral Correction.

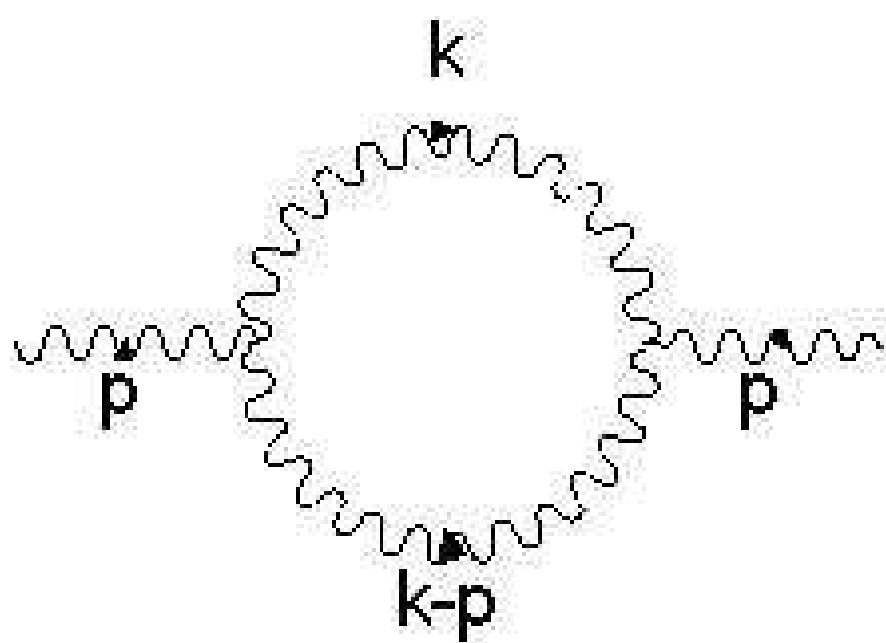
Figure 2: Vector Correction.

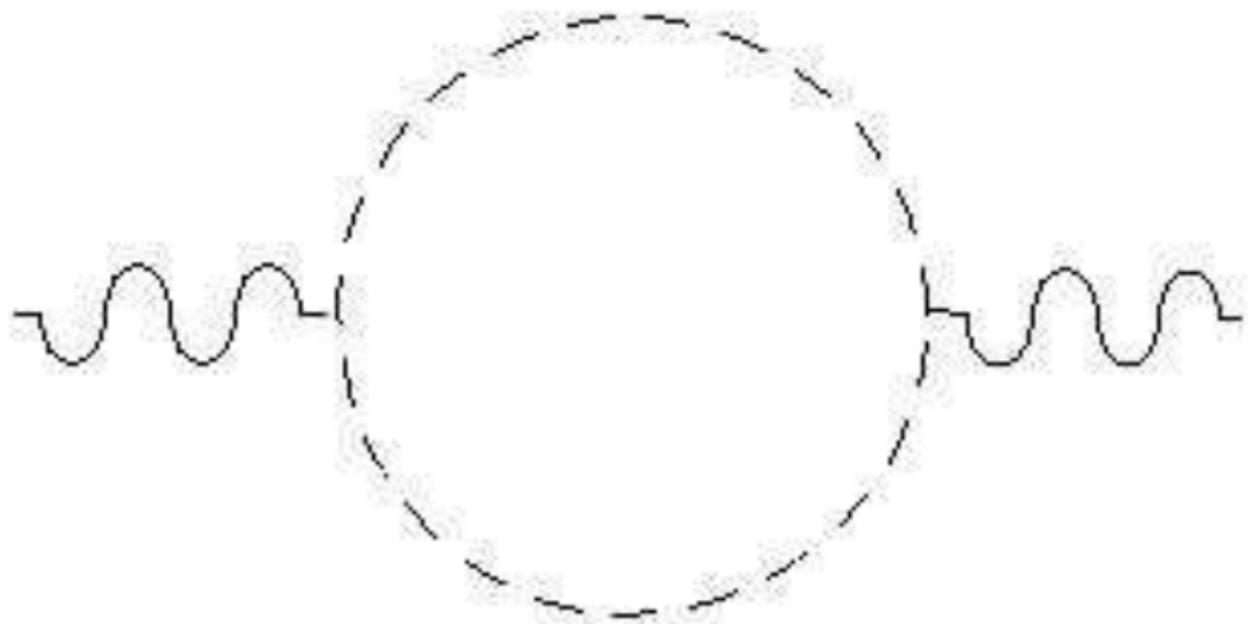
Figure 3: Ghost Correction.

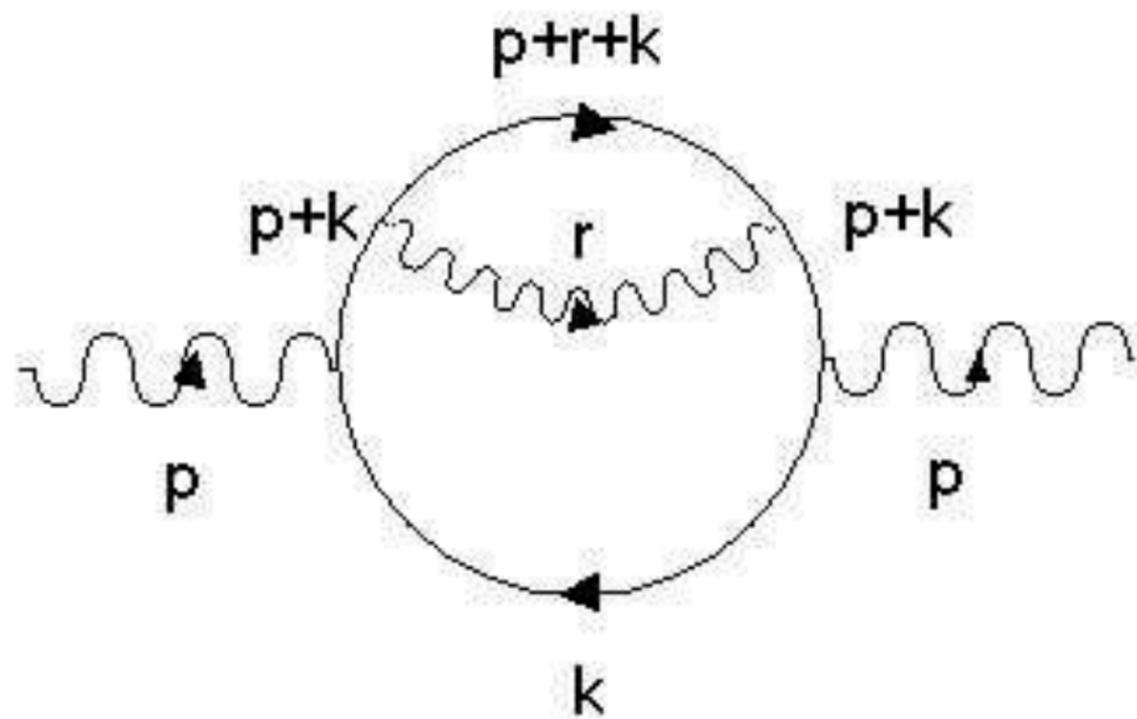
Figure 4: Quantum Correction of Diagram \#4 to the Effective Action.

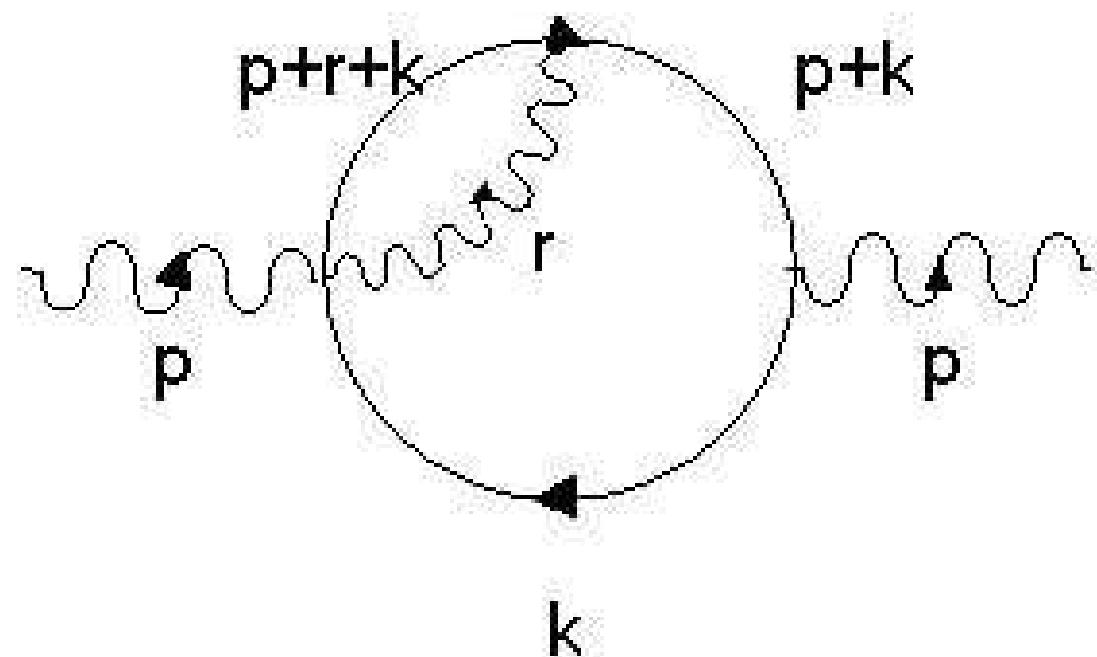
Figure5: Quantum Correction of Diagram \#5 to the Effective Action.

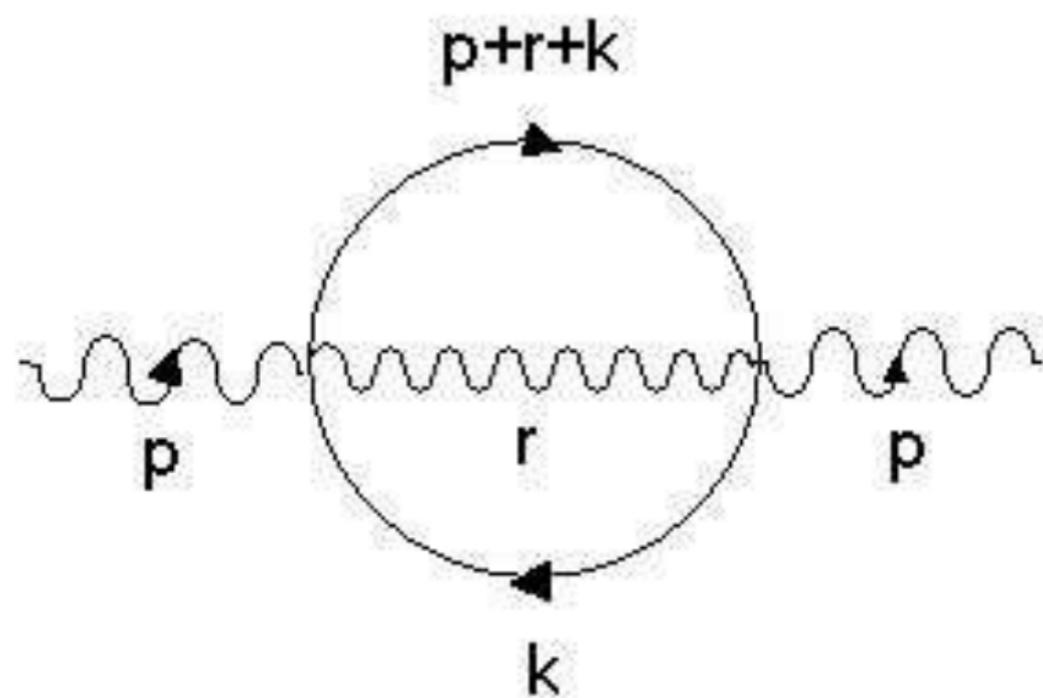
Figure6: Quantum Correction of Diagram \#6 to the Effective Action.

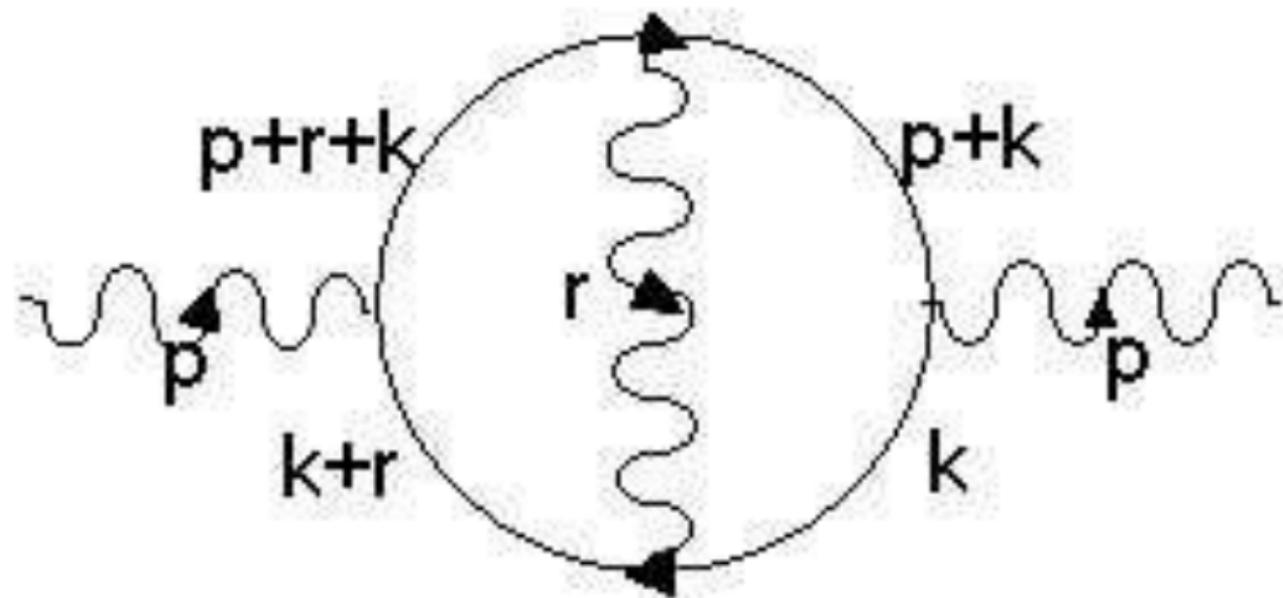
Figure7: Quantum Correction of Diagram \#7 to the Effective Action.

\end{document}